\documentclass[manuscript]{aastex}
\slugcomment{accepted, 25 June 2009}

\shorttitle{Star-Forming Regions in M31}
\shortauthors{Kang et al.}

\begin{document}

\title{An Ultraviolet Study of Star-Forming Regions in M31}

\author{Yongbeom Kang\altaffilmark{1,2}, Luciana Bianchi\altaffilmark{2,3}, and Soo-Chang Rey\altaffilmark{1,3}}
\altaffiltext{1}{Department of Astronomy and Space Science, Chungnam National University, Daejeon, 305-764, Korea; ybkang@cnu.ac.kr, screy@cnu.ac.kr}
%\email{ybkang@cnu.ac.kr}
\altaffiltext{2}{Department of Physics and Astronomy, Johns Hopkins University, Baltimore, MD 21218, USA; bianchi@pha.jhu.edu}
%\email{bianchi@pha.jhu.edu}
\altaffiltext{3}{Corresponding authors}

\begin{abstract}

We present a comprehensive study of star-forming (SF) regions in the nearest large spiral galaxy M31. 
We use $GALEX$ far-UV (1344-1786~\AA, FUV) and near-UV (1771-2831~\AA, NUV) imaging to detect 
young massive stars and trace the recent star formation across the galaxy. 
The FUV and NUV flux measurements of the SF regions, combined with ground-based data for 
estimating the reddening by interstellar dust from the massive stars they contain, are used to derive 
their ages and masses. 
The $GALEX$ imaging, combining deep sensitivity and coverage of the entire galaxy, provides a complete 
picture of the recent star formation in M31 and its variation with environment throughout the galaxy. 
The FUV and NUV measurements are sensitive to detect stellar populations younger than a few hundred Myrs. 
%%We detected and measured 894 SF regions, with size $\ge$1600~pc$^2$ 
We detected 894 SF regions, with size $\ge$ 1600~pc$^2$ 
%above an average flux limit of $\sim$26~mag~arcsecond$^{-2}$, 
above an average FUV flux limit of $\sim$26~ABmag~arcsecond$^{-2}$, 
over the whole 26~kpc (radius) galaxy disk. 
We derive the star-formation history of M31 within this time span. 
The star formation rate (SFR) from the youngest UV sources (age $\le$ 10~Myr) is comparable to 
that derived from H$\alpha$, as expected. 
We show the dependence of the results on the assumed metallicity. 
When star formation detected from IR measurements of the heated dust is added to the UV-measured 
star formation (from the unobscured populations) in the recent few Myrs
%the SFR is slightly decreasing in the recent epochs, 
, we find the SFR has slightly decreased in recent epochs, 
with a possible peak between 10 and 100~Myrs, and 
an average value of SFR $\sim$0.6 or 0.7~M$_{\sun}$~yr$^{-1}$ (for metallicity Z=0.02 or 0.05 
respectively) over the last 400~Myrs.

\end{abstract}


\section{Introduction}

%In the cold dark matter frame, the large spiral galaxies are built on the hierarchical formation scenario. 
%Many observational evidences of galaxies interaction have been acquired in the past decades. 
%However, we still face many challenges in understanding the details of galaxies evolution. 
%Galaxies consist of various stellar populations, interstellar medium (ISM), and dark matter. 
%In particular, young massive stars are a robust tracer of recent galaxy evolution due to 
%their short life. 
In the cold dark matter framework, large spiral galaxies are built hierachically, and there is much 
observational evidence of galaxy interactions to support this. 
%%However, we still face many challenges in understanding the details of galactic structure and evolution. 
However, we still face many challenges in understanding the details of galaxy evolution. 
In this regard, it is interesting to study the hot, massive stellar populations in nearby galaxies 
%%which can be a robust tracer of recent star formation activity. 
which can be a robust tracer of recent star formation activity due to their short lives. 
Young massive stars contribute to the global characteristics at the current epoch of their host 
galaxy and have a major role in the galaxy evolution.

%Young massive stars emit powerful UV radiation. 
%In particular, far-UV imaging is ideal to detect these stars, which are confused with 
%older stellar populations in observations at longer wavelengths \citep[][and references therein]{bia07a}.   
%Most young massive stars are born in associations, therefore we can define these associations 
%from far-UV imaging and study the spatial structure and intensity of the recent star formation. 
%$GALEX$ far-UV (FUV) and near-UV (NUV) bands are ideal to study young associations. 
%The FUV - NUV colour of young stellar populations is very sensitive to the age of the 
%population \citep[e.g.][Fig.~9]{bia08}, because massive stars evolve rapidly. 
%Therefore, UV data allow us to unambiguously identify the youngest stellar populations, to 
%estimate their ages from colours, and the masses from the extinction-corrected UV luminosities. 
%These results are used to study the recent star formation in galaxies, its history and modalities. 
Young massive stars emit powerful ultraviolet (UV) radiation, therefore UV imaging is ideal to detect 
and characterize 
these stars, which are confused with older stellar populations in observations at longer wavelengths 
\citep[][and references therein]{bia07a}.
In most cases, young massive stars are formed in associations in the  galactic disk. 
We define these stellar associations from UV imaging, in order to  study the spatial structure and intensity 
of the recent star formation.
%Thanks to the successful launch and operation of 
{\sl Galaxy Evolution Explorer} \citep[{\sl GALEX},][]{mar05} 
imaging  in far-UV (FUV) and near-UV (NUV) passbands is particularly  useful to study 
massive stellar populations. 
Specifically, the integrated FUV $-$ NUV color of young stellar populations is very sensitive to the age of the 
population, owing  to the rapid evolution of the most massive stars (e.g. Bianchi 2007, 2009:~Fig.~9). 
%\citep[e.g.][Fig.~9]{bia07a,bia09}. 
Consequently, UV imaging data allow us to unambiguously identify the young stellar populations and 
to estimate their ages and masses from colors and extinction-corrected UV luminosities, respectively. 
These results provide the means of probing the history and modality of recent star formation in galaxies.

%We investigate the star-forming (SF) regions in the nearest large spiral galaxy M31. 
%This work presents the first extensive study of the young stellar populations from 
%UV imaging, covering entirely the M31 disk within 26 kpc radius, and extending beyond 
%this radius in some fields. 
%M31 is one of the two largest galaxies in the Local Group, with the Milky Way. 
%It contains a larger number of satellites and globular clusters than our Galaxy. 
%\cite{iba01} found the Giant Southern Stream of M31 and several works 
%found other structures from galaxies interaction in the halo of M31. 
%However, M31 is a quiscent spiral galaxy, having a lower star formation rate (SFR) than our Galaxy. 
%In this work, we study the SF regions in M31 traced by UV imaging and derive the recent star 
%formation history of this galaxy. 
%%M31 is the nearest giant spiral galaxy, only massive galaxies in the Local Group along with the Milky Way. 
M31 is one of the two largest spiral galaxies in the Local Group, along with the Milky Way. 
However, there has been growing evidence that the Milky Way and M31 have different properties 
\citep[][and references therein]{iba07}. 
M31 shows a lower star formation rate (SFR) than the Milky Way \citep{ken98,mas07,hou08,fuc09}.
Furthermore, there are suggestions that M31 appears to be a more typical spiral galaxy than the Milky Way 
\citep{ham07}. 
As with the bulk of local spirals, M31 shows evidences for a formation and evolution history 
affected by merging and/or accretion events, including substructures in its halo \citep[][and references therein]{ham07}. 
In this respect, it is important to investigate the star-forming (SF) regions in M31. 
We present the first extensive study of the young stellar populations from UV imaging, covering 
the entire M31 disk within 26 kpc radius, and extending beyond this radius in some fields. 
%%Studying the SF regions in M31 traced by UV imaging we investigate the recent star formation history of this galaxy.
By studying the SF regions in M31 traced by UV imaging, we investigate the recent star formation history of this galaxy.

In section 2, we describe the {\sl GALEX} UV data used. 
We describe the detection and photometry of SF regions, from UV imaging, in section 3, 
and estimate their interstellar extinction in section 4. 
%We also derive for comparison OB associations from ground-based photometry of stellar sources 
%in section 5. 
%The age and mass of the SF regions are derived in section 6, the star formation history in section 7, 
%and the conclusions in section 8.
%In section 2, we describe {\sl GALEX} UV data for M31. We present the selection and photometry of 
%SF regions of M31, detected from UV imaging, in section 3. 
%In section 4, we estimate interstellar extinction of SF regions. 
In section 5, we select OB associations from ground-based photometry of stellar sources and compare 
them with our UV-defined SF regions. 
The age and mass of the SF regions are derived in section 6, the star formation history in section 7, 
and the conclusions are presented in section 8.

\section{Data}

We focus on the disk region of M31 where most of the recent SF regions are located. 
We adopt the disk semi-major axis value of R = 26~kpc from \cite{wal88}. 
We considered 20 fields from the {\sl GALEX} fourth data release (GR4). 
{\sl GALEX} observed the disk region of M31 as part of the Nearby Galaxies Survey \citep[NGS,][]{bia03}. 
We rejected the fields which have only FUV observations and exposure time shorter than 2,000~s 
(except for ``PS\_M31\_MOS03"), then selected the fields closer to the major axis when different 
pointings are located along the galaxy's major axis. 
The ``PS\_M31\_MOS03" field has a shorter exposure than 2,000~s but it is the only {\sl GALEX} field 
observed in the south outermost disk region of M31. 
As a result, we selected 7 {\sl GALEX} fields covering the entire disk region (see Fig. 1 and Table 1).
Each selected {\sl GALEX} field has FUV (1344-1786~\AA) and NUV (1771-2831~\AA) imaging with the 
same exposure time. 
{\sl GALEX} FUV and NUV imaging has 4.2 and 5.3~arcsec resolution (FWHM) or about $\sim$19~pc in M31, 
and the field of view (FOV) is about 1.27 and 1.25 degree respectively \citep{mor07}. 
We used in our analysis only the central 1.1~degree diameter portion of the field, for best photometry. 
{\sl GALEX} images have a sampling of 1.5~arcsec~pixel$^{-1}$ which corresponds to 5.67~pc, 
%at the distance of M31 \citep[785~kpc,][]{mcc05}. 
%%we assume a distance of 785~kpc \citep{mcc05}. 
assuming a distance of 785~kpc \citep{mcc05}. 

\section{Photometry of Star-Forming Regions}

The FUV images provide a direct measure of the flux from young massive stars not heavily embedded in 
interstellar dust. 
Most SF regions are gravitationally unbound systems 
%and do not have a specific shape. 
and have irregular shapes. 
Rather than using aperture photometry, we constructed contours of the SF regions to trace their 
%%morphology, and measure their UV flux, and flux density. 
morphology, and measure their UV flux and flux density. 
%Basically, we adopt procedure developed by \cite{tol09} for the detection and photometry of SF regions in M31.
%A somewhat similar procedure was developed by \cite{tol09}. 
%We modified this process for more precise detection and photometry of the SF regions. 
The procedure was originally developed by \cite{tol09} in his dissertation, and we have 
modified the procedure for a more precise detection and photometry of the SF regions.
Our procedure consisted of three steps. 
The first step was to detect all image pixels which have flux above a certain threshold in each FUV image. 
An important factor in detecting and defining SF regions is the brightness limit. 
The second step was to define the contours of each SF region from contiguous pixels detected 
above the threshold over a minimum area. 
%%In this way, we can define SF regions contours even if these have a complicated shape, and reject isolated 
In this way, we can define contours of SF regions even if they have a complicated shape, and reject isolated 
stars which are smaller than the minimum size. 
The third step was to estimate the background and to measure the flux of the defined SF regions. 
Even though the background in the UV image is much lower than in the optical, it is important 
to correctly subtract its contribution from the source photometry. 
We performed various tests for determining the optimum flux threshold, minimum size of a SF region, 
and best background subtraction method.

We used the field ``NGA\_M31\_MOS0" which has the longest exposure time (6,811~s) in our selected {\sl GALEX} 
data, to test and refine our procedures, which were then applied to all our selected fields. 
This field is good for testing various types of SF regions because it contains portions from innermost 
to outermost spiral arms and the large OB association NGC 206. 
First of all, we compared various thresholds for the detection of source pixels. 
We used the background-subtracted image (``-intbgsub") provided by the GR4 pipeline. 
We estimated the mean background value by the sigma clipping method. 
We examined the results using thresholds of two, three, and five sigma above the mean background 
value (see Fig. 2). 
We detect fainter objects if we use the lower thresholds, however the SF regions in the spiral arms 
merge together and the contamination by the background (including older, diffuse populations) is larger. 
We can easily define the bright regions if we use higher thresholds, however we cannot detect the faint ones. 
The FUV magnitude limits of our detected SF regions from each threshold are shown in Fig. 3, 
they are $\sim$21.5 (low), $\sim$21.0 (mid), and $\sim$20.4 (high threshold) mag in the AB magnitude system. 
We adopted a threshold of three sigma above the mean background value, which showed in our analysis 
less contamination by background and marginally detects the faint regions. 
This results in an average FUV flux threshold of $\sim$0.0032~c~s$^{-1}$~pixel$^{-1}$, or 
$\sim$25.9~mag~arcsec$^{-2}$. 
Then, we considered the minimum acceptable size of the SF regions, in order to eliminate contamination 
by background objects, artificial sources, foreground stars, and isolated bright stars in M31. 
We considered 3$\times$FWHM of {\sl GALEX} ($\sim$13~arcsec or $\sim$8.5~pixels) as minimum diameter of a SF region, 
therefore we adopted a requirement of a minimum of 50 contiguous pixels ($\sim$1,600~pc$^2$ or $\sim$40~pc) for 
the smallest SF region. 
We contoured the contiguous pixels selected for measuring the defined regions and we estimated the center 
%of each SF region by the maximum diameter of its contour.
of each SF region by the mid-point of the maximum diameter of its contour.

These adopted threshold and minimum size of the SF regions, were then applied to all the selected fields. 
%However, the exposure time differences among fields induce variation of the detection limit. 
However, the exposure time differences among fields induces variation of the detection limit. 
Therefore, we compared the detection fraction in overlapping image regions, one by one against 
the ``NGA\_M31\_MOS0" field. 
%First, we used a fixed threshold ($\sim$0.0032~c~s$^{-1}$~pixel$^{-1}$) for source detection in the ``NGA\_M31\_MOS0" image. 
We first tried a fixed threshold ($\sim$0.0032~c~s$^{-1}$~pixel$^{-1}$) for source detection in the ``NGA\_M31\_MOS0" image. 
In this case, in images with shorter exposure time sources are over-detected and include noise. 
%Second, we adopted the variable threshold of three sigma above the mean background value from each image. 
A second method used the variable threshold of three sigma above the mean background value from each image. 
This produced similar contours in the overlapping regions. 
However, some sources in the shorter exposure time images were undetected because the background has 
larger sigma values. 
%Third, we adopted the ratio between detected and un-detected pixels of overlapping regions. 
Finally, we adopted the ratio between detected and undetected pixels of overlapping regions. 
This case produced slight over-detection in the shorter exposure time images. 
%Therefore, we finally used the third method, manually varying the thresholds in each field, such as 
Therefore, we used the last method, manually varying the thresholds in each field, such as 
to obtain similar detections in overlapping regions. 
The final selected FUV pixel maps are shown in Fig. 4. 
%As last step, we compared the detected SF regions within overlapping image regions by visual inspection 
As a last step, we compared the detected SF regions within overlapping image portions by visual inspection 
to select a final catalog of unique sources. 
%In the end, we obtained a catalog of 894 SF regions clean of artificial objects, isolated stars, 
At the end, we obtained a catalog of 894 SF regions cleaned of artifacts, isolated stars, 
and overlapping objects. 

We measured the flux within the contours of the SF regions as defined above, and subtracted the local background, 
estimated in a circular annulus surrounding the source. 
We initially used an inner/outer size of the annulus of 1.5/3 times the size of each SF region. 
However, for the largest sources, which are mostly found along the spiral arms, 
the annulus for background measurement scaled in this way becomes too large and includes unrelated 
stellar populations. 
Therefore, we tested three different procedures to estimate the background. 
One was to measure the local background (``Ap. sky" in Fig. 5), adjusting the radius of the background annulus 
according to the size of the source. 
We adopted a variable scaling of inner/outer radii for the background annulus, adjusted as 
2/4, 2/3, 1.5/2.5, 1.1/1.3, and 1.05/1.2 times the source size (R$_{MAX}$) for the following ranges of 
source size respectively: R$_{MAX}$ $<$ 10, 10 $\le$ R$_{MAX}$ $<$ 20, 20 $\le$ R$_{MAX}$ $<$ 50, 
%%50 $\le$ R$_{MAX}$ $<$ 100, 100 $\le$ R$_{MAX}$ ([pixels]). 
50 $\le$ R$_{MAX}$ $<$ 100, and 100 $\le$ R$_{MAX}$ ([pixels]). 
%Such range was found adequate to ensure a large enough area for the background measurement for 
Such ranges were found adequate to ensure a large enough area for the background measurement for 
the smallest sources, while preventing excessively large areas to be included in the calculations for 
large sources. 
The second background estimate was obtained from the GR4 pipeline background image (``-skybg") 
, and the third measurement was performed by applying a median filter to the image. 
In the case of the GR4 pipeline background (``Pipe. sky" in Fig. 5), a sky background image is produced by 
a 5$\times$5 median filter size which is good for the case of Poisson distribution \citep{mor07}. 
The background measured using this image, underestimates the local contribution by the galaxy's background light, 
especially from surrounding stellar populations in the spiral arms. 
In the case of the median filtered background (``Med. sky" in Fig. 5), a smaller filter size (3$\times$3 pixels) 
%%than the standard pipline was used to produce the background images. 
than the standard pipeline was used to produce the background images. 
The choice of a smaller filter size was driven by the consideration of spiral arm regions which host 
most of the SF regions. 
The results from this background estimate are similar to those from the local background and reflect well the 
brightness of spiral arms. 
The results from the three methods are compared in Fig. 5. 
The background from the pipeline 'sky' image is always underestimated because it measures the lowest sky 
level and not the local diffuse stellar population surrounding the source. 
In the right-side panels in Fig. 5, we see that this estimate is not sensitive to the spiral arm 
enhancements, as the local-background estimates are. 
The local sky estimate is similar to the results from the median-filtered measurements but 
shows less scatter for sources with multiple observations in overlapping regions. 
In Fig. 6, we compare the photometry results using the three different methods for background subtraction 
%%as color-magnitude diagrams. 
in color-magnitude diagrams. 
Most SF regions have FUV $-$ NUV color between $-0.5$ and $1.0$ in ABmag. 
The measurements from the pipeline and median-filtered backgrounds induce brighter NUV than FUV, especially 
for sources fainter than 20~mag. 
We finally adopted the result from the local background. 
The resulting catalog of SF regions and their {\sl GALEX} photometry is given in Table 2.

\section{OB Stars and Interstellar Extinction}

In order to derive the physical parameters of the SF regions from the integrated photometry, we must 
take into account the interstellar extinction. 
We estimated the reddening of each UV SF region from the reddening of the massive stars included within 
its contour. 
For massive stars we used the reddening-free parameter Q \citep{mas95,bia06}. 
We used the optical point-source measurements of M31 sources from P. Massey (priv. comm., 2009), which is 
a revised M31 catalog from the NOAO survey data described by \cite{mas06, mas07}. 
We selected sources with $UBV$ measurements having photometric errors lower than 0.1~mag in all bands 
(108,089 objects out of their 371,781 total sources catalog), which results in a magnitude limit of 
about 23rd~mag in $V$-band. 
This data is deep enough to select OB type stars which we used to estimate the interstellar reddening 
of our SF regions. 
We selected OB type stars by comparing colors and brightness of the Galactic OB type main sequence 
stars from \cite{all82}. 
We selected stars from O3 to B2V, because a B2V star has $M_V$ = $-2.45$ which is 
$V$ = $22.02$ in M31 ($m - M$ = 24.47) in absence of extinction. 
We don't know the internal reddening of M31, therefore we used the reddening-free parameter 
$Q_{UBV}$ to select the OB stars. 
\begin{center}
$Q_{UBV}$ = $(U-B)$ - ${E(U-B)}\over{E(B-V)}$$(B-V)$
\end{center}
The $E(U-B)/E(B-V)$ ratio is a constant (0.72) for Milky Way dust type \citep{mas95} and does not vary 
much for other dust types, within a reasonably small range of $E(B-V)$ 
\citep[e.g.][and references therein]{bia07,bia06}. 
%%%@@@ check below !! 
We selected the OB stars which have $Q_{UBV}$ between $-0.97$ (O3V) and $-0.67$ (B2V). 
%In order to reduce contamination, we also adopted magnitude and color limits; -6.0 $\le$ $M_V$ $\le$ -0.9, 
%-0.34 $\le$ $B-V$ $\le$ 0.26, and -1.22 $\le$ $U-B$ $\le$ -0.48 \citep{all82}.
In order to reduce contamination, we also adopted magnitude and color limits; $-6.0$ $\le$ $M_V$ $\le$ $-0.28$, 
$-0.34$ $\le$ $B-V$ $\le$ 0.46, and $-1.22$ $\le$ $U-B$ $\le$ $-0.336$ \citep{all82}.
 The color and magnitude limits are derived assuming a maximum reddening value of $E(B-V)$ $\le$ 0.7.
% minimum \citep{sch98} and maximum \citep{mag93} 
%reddening values (0.06 $\le$ $E(B-V)$ $\le$ 2.0). 
%The location of the selected OB stars in color-magnitude, color-color, and color-Q diagrams are presented 
The location of the selected OB stars in color-magnitude diagrams are presented 
in Fig. 7.
With these restrictions, we finally selected 22,655 O-B2 stars from the data of \cite{mas06}. 
The spatial distribution of these stars, shown in Fig. 1, represents well the spiral structure of M31. 

The interstellar extinction of the selected OB stars was estimated from the reddening free parameter Q$_{UBV}$, 
using the empirical relationship for giant and main-sequence stars by \cite{mas95}. 
\begin{center} 
$(B - V)_0 = -0.013 + 0.325Q_{UBV} = (B - V) - E(B - V)$ 
\end{center}
The estimated $E(B-V)$ has mean, median, and mode value of about 0.34, 0.32, and 0.29, respectively, from
our selected  OB stars.
%%%@@@ maximum value of about 1.2 and a 
%%%@@@ check number 
%%
%%The mean reddening value of OB stars is bigger than \cite{mas07}'s mean value $E(B-V)$ = 0.13 which estimated from 
The mean reddening value of OB stars is larger than \cite{mas07}'s typical value $E(B-V)$ = 0.13 which is 
estimated visually from the location of the ``blue plume'' in the color-magnitude diagram.
%blue supergiants. 
%%This difference is caused by selection of blue stars for estimating reddening. 
This difference may be caused by our selection of blue stars for the  reddening estimate. 
We also explore in this paper more reddened regions than the average entire stellar population.  
%therefore selected stars from O3V to B2V and assumed 
%the maximum reddening 2.0. 
We estimated the extinction by interstellar dust for each SF region from the average reddening of 
the OB stars within the SF region contour. 
For the SF regions outside of 
%In the cases where there are no OB stars in the SF contour, because they are outside of 
\cite{mas06} survey, we assumed $E(B-V)$ = 0.20. 
The spatial distributions of estimated interstellar reddening of OB stars and SF regions are shown in Fig. 8. 
The interstellar extinction decreases from the inner disk region outwards. 
In particular, inner-most and south-west areas where we could not detect SF regions, have high interstellar 
extinction. We will return to this point in the next section. 
Our detection method is based on the FUV flux, which could vanish in high interstellar extinction regions. 

%%\section{OB associations defined from stellar photometry}
\section{OB Associations Defined from Stellar Photometry}

For comparison with our FUV-defined SF regions, we also used a Path Linkage Criterion 
\citep[PLC:][]{bat91} method \citep[explored by][]{iva96,iva98,mag93,tol09} to detect OB associations 
using O-B2 stars. 
We applied the PLC method varying the minimum number of stars (N$_{min}$) and maximum link 
distance (d$_{s}$) (see Fig. 9). 
The best choice of N$_{min}$ and d$_{s}$ was found to be N$_{min}$ = 5 stars and d$_{s}$ = 10.4 
arcsec ($\sim$40~pc). 
\cite{mag93} found 174 OB associations and estimated a total number of $\sim$420 associations in M31 
from a similar method but different optical photometry. 
With this method, we found 650 OB associations in M31 from the O-B2 stars selected by us from 
\cite{mas06} photometry, which is $\sim$ 2 mag deeper than what \cite{mag93} used. 
We compared these OB associations with the 894 SF regions selected from FUV imaging. 
They mostly overlap with the FUV-detected regions (see Fig. 10), however some of FUV-selected SF regions 
have a larger area than OB associations defined from stellar photometry, and some additional 
OB associations are found from stellar photometry in high interstellar extinction regions. 
The observation fields of \cite{mas06} cover a smaller area than our {\sl GALEX} imaging, therefore we compared 
our results within 17~kpc de-projected distance (about 1~deg$^2$; dashed ellipse in Fig. 10) from 
the center of M31. 
%%The total area of the SF regions derived from the FUV contours (section 3), and of the 
The total area of the SF regions derived from the FUV contours (Section 3), and of the 
%OB associations derived from stellar photometry (above), is a fraction of 0.041 and 0.037 of 
OB associations derived from stellar photometry (above), are 4.1~\% and 3.5~\% of 
the area of the 17~kpc disk, respectively.
%%%@@@ check numbers below  
The numbers of O-B2 stars are 9094/11350 and 8670/11774 inside/outside of the UV-defined SF regions 
and of OB associations. 
%%The average projected density of OB stars in the associations is 0.015 and 0.017~stars~arcsec$^2$ from UV-selected 
The average projected density of OB stars in the associations is 0.017 and 0.018~stars~arcsec$^{-2}$ from UV-selected 
SF regions and OB associations selected from stellar photometry, respectively. The projected density
of stars inside SF regions is  about a factor of 20
higher than in general field. 
The comparison between the two methods is interesting, because the FUV-selected contours are more affected 
by interstellar reddening than the optical stellar photometry, on the other hand optical bands are not as 
sensitive as the FUV is to the T$_{eff}$ of the hottest stars \citep[e.g.][and references therein]{bia07a}. 
The similarity of the total number of OB stars detected inside young associations and in the field, by the two 
methods is remarkable. 
The slight difference in the estimated area of the associations may be due, at least partly, 
to the low ($\approx$5~arcsec) spatial resolution of the {\sl GALEX} imaging.

\section{Ages and Masses of Star-Forming Region}

We estimated the ages of the SF regions by comparing the measured (FUV $-$ NUV) colour with synthetic 
Simple Stellar Population (SSP) models, reddened by the extinction amount estimated for each region. 
We explored effects of metallicity and dust type on the results 
\citep[e.g.][]{bia07a,bia09}. 
Then we estimated the masses of the SF regions from the reddening-corrected UV luminosity and the derived ages. 
We used two sets of SSP models, one from \cite{bru03} (BC03) and the other from Padua 
(PD: A. Bressan, priv. comm., 2007). 
The ages derived from the two grids of models do not differ significantly (see lower left panel of Fig. 11). 
Ages estimated using the BC03 models tend to be slightly younger than those derived using the PD models, 
below 30~Myrs, and do not differ at all for older populations (Fig. 11, lower left panel). 
The small age difference propagates to the derived masses, as shown in the lower-right panel of Fig. 11. 
The differences between results from the two model grids is not significant. 
We used the PD models in our analysis. 

%%Most SF regions have UV colour between -0.5 and 1.0 (see the colour-magnitude diagram in Fig. 11). 
Most SF regions have UV colour between $-0.5$ and 1.0 (see upper left panel of Fig. 11). 
The estimated ages of most SF regions in our sample are younger than 400~Myrs, reflecting our 
FUV-based selection. 
Our detection limit, plotted with a line in Fig. 11, indicates quantitatively 
how the flux-detection limit translates into mass detection limit, as a function of age, 
and shows that we cannot detect low mass SF regions at older ages, as expected. 
However, we also notice a lack of massive SF regions at younger ages. 
This will be discussed later.

%%We explored three metallicity values: subsolar (Z=0.008), solar (Z=0.02) and supersolar (Z=0.05) metallicity, 
We explored three metallicity values: subsolar (Z=0.008), solar (Z=0.02), and supersolar (Z=0.05) metallicity, 
although M31 is believed to have a typical metallicity about twice higher than the MW 
\citep[e.g.][and references therein]{mas03}. 
Our census of young stellar populations based on wide-field FUV imaging has an unprecedented extent, 
while direct metallicity measurements from spectroscopy are confined to limited samples. 
%%Therefore, we wanted to assess in general the dependence on our results on metallicity, which may vary in some 
Therefore, we wanted to assess in general the dependence of our results on metallicity, which may vary in some 
environments. 
We also examined the effect of four types of interstellar dust: Milky Way ($R_V$ = 3.1; MW, \cite{car89}), 
average Large Magellanic Cloud (AvgLMC), 30 Doradus (LMC2) \citep{mis99}, 
and Small Magellanic Cloud \citep[SMC,][]{gor98} dust extinction. 
The resulting ages and masses of the SF regions are plotted in Figs 11, 12, and 13. 

The differences in derived ages and masses for three metallicity values and different types of interstellar 
dust are presented in Fig. 12. 
We considered metallicity values of no less than Z=0.008 because we expect the young SF regions not 
to be as metal poor as old globular clusters. 
In Fig. 12 ages and masses derived from models with solar metallicity, and assuming MW-type interstellar 
reddening (R$_V$ = 3.1), are compared to results from subsolar (Z=0.008) and supersolar (Z=0.05) 
%%metallicities (top panels). 
metallicities (top four panels). 
%The derived ages are older for subsolar metallicity, and younger for supersolar metallicity, respect 
The derived ages are older for subsolar metallicity, and younger for supersolar metallicity, with respect 
to solar metallicity results. 
The differences are most significant for ages younger than 100~Myrs \citep[see also Fig.~9 of][]{bia09} 
and are up to a factor of $\sim$3 at most. 
Because the effect is stronger at certain ages, the number distribution of SF regions with ages also 
differs, as shown in Fig. 12. 
This will be taken into account in the following analysis, where we derive the global SF in M31 
as a function of time. 
The difference in the derived masses is not conspicuous, considering the uncertainties. 
The uncertainty of the ages derived by comparing the photometry to synthetic population models, 
reported in Table 2, is derived by propagating only the photometric errors, because we investigated and showed 
separately the effects of different metallicity values and dust types. 
The uncertainty on the derived masses reflects the uncertainty on the photometry and the age. 
Reddening corrections, as derived in Section 5, are applied. 

%%The lower panels of Fig. 12 show the effects of the correction for reddening. 
The lower four panels of Fig. 12 show the effects of the correction for reddening. 
If the selective extinction by dust is steeper in the UV than the MW dust, as observed 
for example in the LMC (average) or the extreme starburst regions 30 Dor (labelled as ``AvgLMC" and ``LMC2", 
respectively, in Fig. 12), the dereddened UV luminosity will be higher but the dereddened FUV $-$ NUV 
color bluer, resulting in much younger ages and consequently lower masses. 
An LMC-type dust is however not likely in M31. 
\cite{bia96} report UV extinction curves in M31 similar to the average MW extinction, from UV spectra 
of OB stars. 
Moreover, if we apply UV extinction steeper than MW dust, most measured FUV $-$ NUV colors 
appear to be over-corrected, when compared to SSP model predictions. 
Out of 847 SF regions whose (FUV $-$ NUV)$_0$ is within the model color range when dereddened with 
MW dust type, only 569/227/26 SF regions have (FUV $-$ NUV)$_0$ colors within the model range if 
the progressively steeper dust types AvgLMC/LMC2/SMC are applied. 
Therefore, UV-steep dust extinction seems to be not realistic in most cases. 

\section{Results}

\subsection{Spatial distribution of the Star-Forming regions}

The FUV-selected SF regions follow the disk structure of M31 and their spatial distribution traces 
the recent star formation in M31 (Fig. 13). 
%%As can be seen in Fig. 13, most SF regions have de-projected distances between 40 and 70~arcmin 
As can be seen in Fig. 13, most SF regions have de-projected distances between 40 and 75~arcmin 
%%(9 and 16~kpc) from the center of M31, with two peaks in this region. 
(9 and 17~kpc) from the center of M31, with two peaks in this region. 
This region is well-known as the ring of fire or the star-formation ring \citep[][and references therein]{blo06}. 
%One thing of interest is that the number of old SF regions (age $>$ 100~Myrs) decreases faster than 
%the young SF regions (age $<$ 100~Myrs) outside of this ring ($d_{de-projected}$ $>$ 70~arcmin). 
%In addition, the outermost disk ($d_{de-projected}$ $>$ 100~arcmin) has only young SF regions and no 
%old SF regions. 
%Young SF regions are spread throughout most of M31's disk. 
%This means that the M31's disk formed stars continuously during the last few hundreds Myrs at least.
%%One thing of interest is that the number (84) of younger SF regions (age $>$ 10~Myrs) is larger than the number (40) of SF regions 
%%which have older than 10~Myrs outside of this ring ($d_{de-projected}$ $>$ 75~arcmin). 
%%The numbers are similar each SF regions which are 300 (younger) and 287 (older) between 40 and 75~arcmin.
%%The number (17) of younger SF regions, however, is less than the number (86) of olders inside of this region. 
%%These reflect that the outer disk more vigorous star formed than inside recently and the region of 
%%star-formation ring formed stars continuously during the last few hundreds Myrs at least.
In this star formation ring between 40 and 75 arcmin from the galaxy center, the number of young ($<$ 10~Myrs) SF regions is similar 
to that of older ($>$ 10~Myrs) ones. 
One thing of interest is that the number  of younger SF regions (82) is larger than the number of older SF regions (40) 
outside of this ring ($d_{de-projected}$ $>$ 75~arcmin). 
Inside of the star formation ring, the number  of younger SF regions (14), however, is less than that  of older ones (89).
This suggests that the M31 disk formed stars continuously during the last few hundreds Myrs at least and, 
furthermore, the outer disk shows more recent star formation.

%%The number ratio (290/297) of less/more masses is similar within the star-formation ring 
%%if the mass separated 5 $\times$ 10$^{3}$~M$_{\sun}$. 
%%The number ratios of less/more massive are 20/83 inside and 87/37 ouside of the ring. 
%%The ratios of smaller/larger size are 48/55, 300/287, and 66/58 from inside to ouside which 
%%separated by 2.8 $\times$ 10$^{3}$~pc$^{2}$. 
%%The raios of less/more dense (mass-per-size) are 13/90, 299/288, and 91/33 which separated by 
%%1.5~M$_{\sun}$/pc$^{2}$. 
The size of some younger SF regions is larger than that of older SF regions, however most of them are 
less massive (as derived from the UV flux) than older regions. 
Young, large SF regions may be broken into several SF regions with time, while some of the dense, small 
SF regions may survive longer than the others. 
%%Therefore, we suggest that the younger and less dense SF regions are more formed outside disk than inside disk of M31.

\subsection{Recent Star Formation History in M31}

We noticed a lack of massive SF regions younger than 50~Myrs in Fig. 11. 
We would expect to find some massive young SF regions if star formation was constant. 
We estimated the total SFR in M31 using the ages and masses of the SF regions 
derived in Section 6. 
We added the estimated initial masses of the SF regions separated in four time bins, to investigate 
the SFR time evolution. 
The results are shown in Fig. 14, and reported in Table 3, for three metallicity values. 
%%!!!@@@ new 
Although in each age interval the derived SFR depends on the assumed metallicity, 
in all cases Fig.14 shows an apparent decrease of SFR in the recent epoch ($<$ 10~Myrs) 
with respect to the average value in the interval 10-100~Myrs ago, 
the difference being smallest (and probably not significant) for supersolar metallicity, 
which is currently believed to be the most probable value for M31. 
%%!!!@@@ old:
%Fig. 14 shows an apparent decrease of the SFR in recent epochs when we assume solar or Z=0.008 metallicities. 
The dashed line across the whole time interval is the average from the SF regions of all ages. 
When interpreting this diagram, we must first of all recall that the masses are estimated 
from the UV flux above a certain threshold, and the corresponding limit for mass detection 
increases at older ages (Fig. 11, continuous line). 
Therefore, the total stellar mass formed over 100~Myrs ago may be relatively underestimated 
compared to that of younger populations, due to our FUV selection. 
This bias makes the apparent decrease in SFR at young ages more robust. 
Most of the time-binned and mean SFRs are lower than one solar mass per year. 
These values can be compared to the M31's SFRs from IR and H$\alpha$. 
\cite{bar06} estimated a SFR of 0.4~$M_{\sun}$~yr$^{-1}$ from 8~$\micron$m non-stellar emission, 
higher than our UV-derived SFR in the $<$4~Myrs bin, if we assume solar metallicity, 
but comparable to our value for supersolar metallicity. 
This shows that each indicator alone, UV or IR, may miss about half of the most recently formed stellar mass. 
\cite{mas07} estimated 0.05~$M_{\sun}$~yr$^{-1}$ from H$\alpha$ luminosities, lower 
(within a factor of two) than our estimate for the $<$10~Myrs bin (solar metallicity). 

\section{Conclusion and Discussion}

We used 7 {\sl GALEX} fields covering the entire disk of M31 (out to $\sim$ 26~kpc radius) to study its young 
stellar population. 
We detected 894 SF regions from the FUV imaging, and measured their integrated FUV and NUV fluxes. 
We estimated the interstellar extinction in each SF region from the OB stars within its contour, 
using the ground-based stellar photometry of \cite{mas06}. 
We estimated ages and masses of our SF regions in M31, detected from FUV imaging, by 
comparing the FUV and NUV measurements to population synthesis models. 
Most are younger than 400~Myrs (UV is not sensitive to older ages). 
%Interestingly, there are no massive SF regions at young ages (Fig. 11), suggesting an apparent 
Interestingly, there are no massive SF regions at young ages (age $<$ 50~Myrs in Fig. 11), suggesting an apparent 
decrease of the SFR, as detected from the FUV imaging.

There may be a potential bias, due to our definition of the SF contours from the FUV image, 
concerning the size distribution and average surface brightness. 
Younger SF complexes tend to be more compact, and older stellar associations are less dense, in general 
\citep{bia07a,efr09}. 
Therefore, the same flux detection algorithm may break a young SF area into several peaks, but these 
contiguous regions may appear connected if they spread out at older ages. 
%This bias, however, does not appear significant at a visual inspection of the complex SF areas along spiral arms, 
This bias, however, does not appear significant from visual inspection of the complex SF areas along spiral arms, 
and younger associations tend to have larger size in our selection. 
In any case, the bias would only affect the apparent distribution [with age] of sizes and masses of 
the individual SF regions, and not affect the results when we add all individual masses in a wide age 
interval, in order to derive the total SFR in M31. 
Because we have estimated ages and masses of the individual SF complexes, and the UV flux is sensitive to 
a much broader age range than e.g. H$\alpha$ or IR, we were also able to derive the recent SFR in M31 as 
a function of time. 
We estimated the SFR in four time intervals adding the masses of the SF regions of corresponding ages. 
The resulting masses are restricted by our detection limit which is shown in Fig. 11. 
The plot in Fig. 14 shows a recent apparent decrease of the SFR, as derived from the UV flux. 
However, we know that the youngest and compact SF regions are often still embedded in dust 
\citep[e.g.][and references therein]{bia07a}. 
These escape detection from UV imaging, but are instead revealed by IR emission from heated dust. 
%%Therefore, for a complete account of star formation in the very young age bin ($<$ few~Myrs), 
Therefore, for a complete account of star formation in the very young age bins ($<$ 10~Myrs), 
we added the IR-measured star formation from \cite{bar06}. 
%%The total SFR at young ages (UV + IR estimates) is shown with an arrow in Fig. 14. 
The total SFR at young ages (UV + IR estimates) is shown as thick lines at the top of vertical arrows in Fig. 14. 
This more realistic and complete estimate of the star formation in the recent few million years significantly 
reduces the apparent recent decrease in SFR, as derived from UV flux only. 

We also show the average stellar mass formed in a $\le$ 10~Myrs bin, from our UV measurements, 
for comparison to H$\alpha$ estimates. 
Because H$\alpha$ is an indirect measurement of the ionizing photons from the short-lived O-type 
stars, the SFR derived from this method must be compared to our age bin of $\le$ 10~Myrs. 
The UV and H$\alpha$ SFR estimates agree within a factor of two, for solar metallicity, as shown in Fig. 14. 
If we assume supersolar metallicity for all sources (Z=0.05), then the H$\alpha$ underestimates 
the SFR by a factor of ten with respect to the SFR derived from the UV sources in the recent 10~Myrs. 
We note that the results shown in Fig. 14 are derived by dereddening the UV fluxes with MW-type dust extinction. 
If UV-steeper reddening applies to the intrinsic dust extinction in some SF regions, their derived ages 
would be younger (see Fig. 12), resulting in a SFR lower for older ages and higher for younger ages 
%%than the values shown in Fig.14, and increasing the discreapancy with the Ha estimate for supersolar metallicity.
than the values shown in Fig.14, and increasing the discreapancy with the H$\alpha$ estimate.
% for supersolar metallicity.

Even when the IR- and UV-derived SFRs are added at the youngest epochs, 
there seem to be a recent decrease of SFR, or a peak of SFR between $\sim$ 10-100~Myrs 
(although the UV estimate of SF at older ages is a lower limit, as previously explained). 
This seems to suggest that M31 had a starburst during this interval. 
%%A possible scenario of galaxies interaction has been discussed that may have induced this violent star formation 
%%around this time interval. 
A possible scenario is that galaxy interaction may have induced violent star formation around this time interval. 
\cite{gor06} considered a collisional event with M32 around 20~Myrs ago to explain the ring 
%%structure by dynamical models, and \cite{blo06} also postulated a head-on collision event with M32 
structure by dynamical models.
\cite{blo06} also postulated a head-on collision event with M32 
and estimated it happened around 210~Myrs ago. 
We now have a possible evidence of a recent starburst in M31, which may have constructed the ring structure.

This work presented the first estimate of the recent SFR based on measurements of individual SF regions 
across the entire galaxy, from UV imaging. 
\cite{mas07} provide a comparison of SF, based on H$\alpha$ measurements, among Local Group galaxies, and several 
%%other authors give estimates of SF in nearby galaxies, however their results are mostly derived under the 
other authors give estimates of SF in nearby galaxies. 
However, their results are mostly derived under the 
%%assumption of continuous star formation (CSP), translating global fluxes into SF. 
assumption of continuous star formation, translating global fluxes into SF. 
Our study provides a time-resolved SFR over the past few hundred million years. 
We will perform a similar analysis on other Local Group galaxies, for a consistent comparison of 
results from our method within a range of physical environments. 
Such comparison should also clarify the relative calibration of UV, IR, and H$\alpha$ as SF indicators 
as a function of galaxy physical conditions. 

In future works, we will also compare the parameters describing the properties of SF regions 
from the integrated measurements with resolved studies of their stellar populations 
(from ground-based and Bianchi's HST programs data). 
We will estimate ages and masses from deeper {\sl GALEX} images to test the limit of our current 
detection threshold.

\acknowledgments
YBK was supported by the Korea Research Foundation Grant funded by the Korean Government(MOEHRD) (KRF-2007-612-C00047). 
SCR acknowledges support from the KOSEF through the Astrophysical Research Center for the Structure and Evolution of the Cosmos (ARCSEC). 
We are grateful to D. Thilker for discussions about the background subtraction, to K. Kuntz for a careful reading of the manuscript and 
useful comments, and A. Tolea for providing some of the procedures he developed for his dissertation with Luciana Bianchi and K. Kuntz. 
We are also grateful to P. Massey for discussions about the reddening value and for providing the revised photometry catalog. 
This work is based on archival data from the NASA Galaxy Evolution Explorer (GALEX) which is operated for NASA by the California Institute of Technology under NASA contract NAS5-98034. 
The GALEX data presented in this paper were obtained from the Multimission Archive at the Space Telescope Science Institute (MAST). 
Support for MAST for non-HST data is provided by the NASA Office of Space Science via grant NAG5-7584 and by other grants and contracts. 
GALEX (Galaxy Evolution Explorer) is a NASA Small Explorer, launched in April 2003. 
We gratefully acknowledge NASA's support for construction, operation, and science analysis of the GALEX mission, 
developed in cooperation with the Centre National d'Etudes Spatiales of France and the Korean Ministry of Science and Technology.

%%%%%%%%%%%Figure%%%%%%%%%%%%%%%

\begin{figure} 
\begin{center}
\includegraphics[width=146mm]{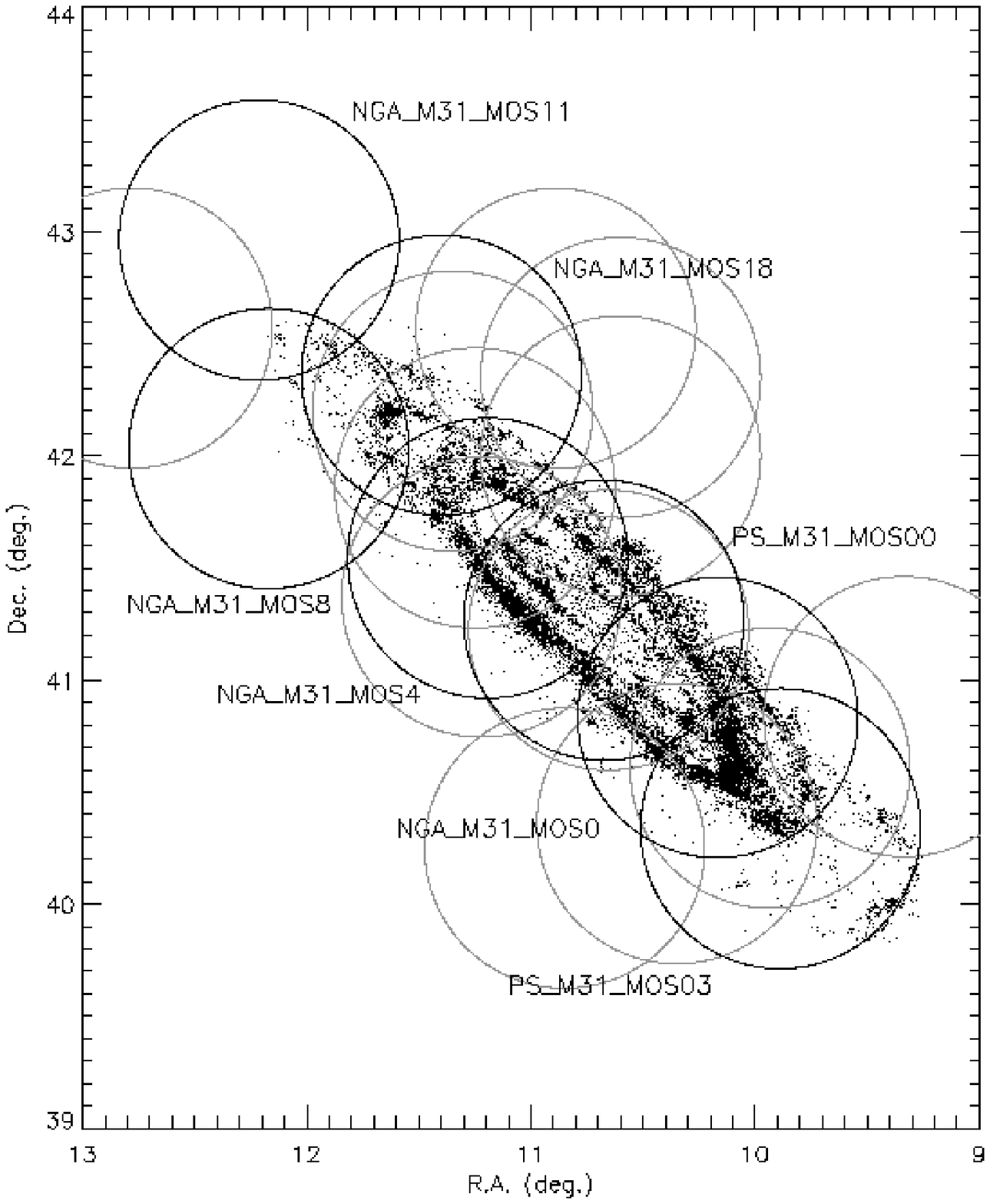} 
\end{center}
\caption{$GALEX$ fields covering the M31 disk region from the GR4 data release. 
Black circles are our selected $GALEX$ fields and grey circles are additional $GALEX$ observations, not selected. 
Black dots are OB type stars selected from the ground-based photometric catalog of \cite{mas06}, 
as described in Section 4.\label{fig1}}
\end{figure}

\begin{figure} 
\begin{center}
\includegraphics[width=166mm]{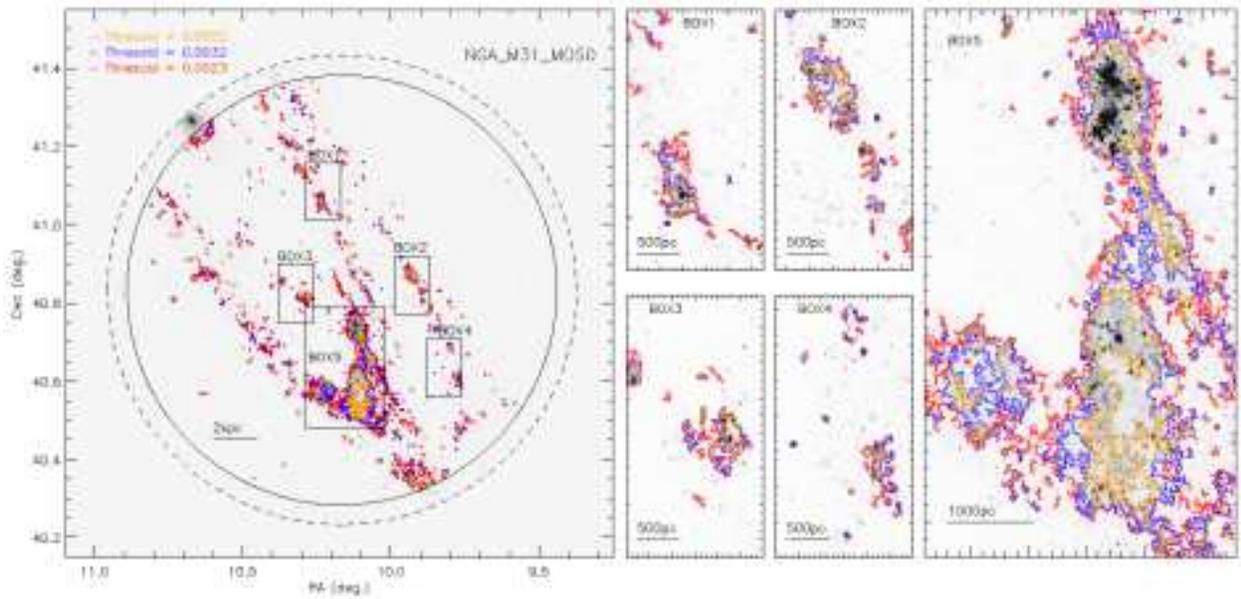} 
\end{center}
\caption{Differences of source detection in the ``NGA\_M31\_MOS0" field using three different flux thresholds. 
Red contours are derived using the low threshold (mean background + 2$\sigma$), blue contours using 
the mid threshold (mean background + 3$\sigma$), and orange contours using the high threshold (mean 
background + 5$\sigma$). 
The solid circle is the FOV used in our analysis and the dashed circle is the whole $GALEX$ FOV. 
Enlargements of sample regions are shown in the rectangular panels. 
Background images are the grey scaled FUV images of the ``NGA\_M31\_MOS0" field.\label{fig2}}
\end{figure}

\begin{figure} 
\begin{center}
\includegraphics[width=83mm]{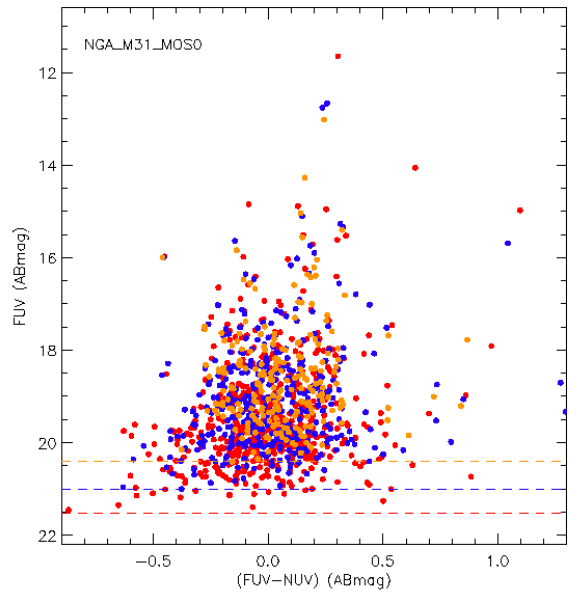}
\end{center}
\caption{The color-magnitude diagram of SF regions in the ``NGA\_M31\_MOS0" field. 
Red dots are detections using the low flux threshold (mean background + 2$\sigma$), blue dots are for 
the mid threshold (mean background + 3$\sigma$), and orange dots are for the high threshold 
(mean background + 5$\sigma$). 
Dashed lines are the magnitude limit of each threshold.\label{fig3}}
\end{figure}

\begin{figure} 
\begin{center}
\includegraphics[width=166mm]{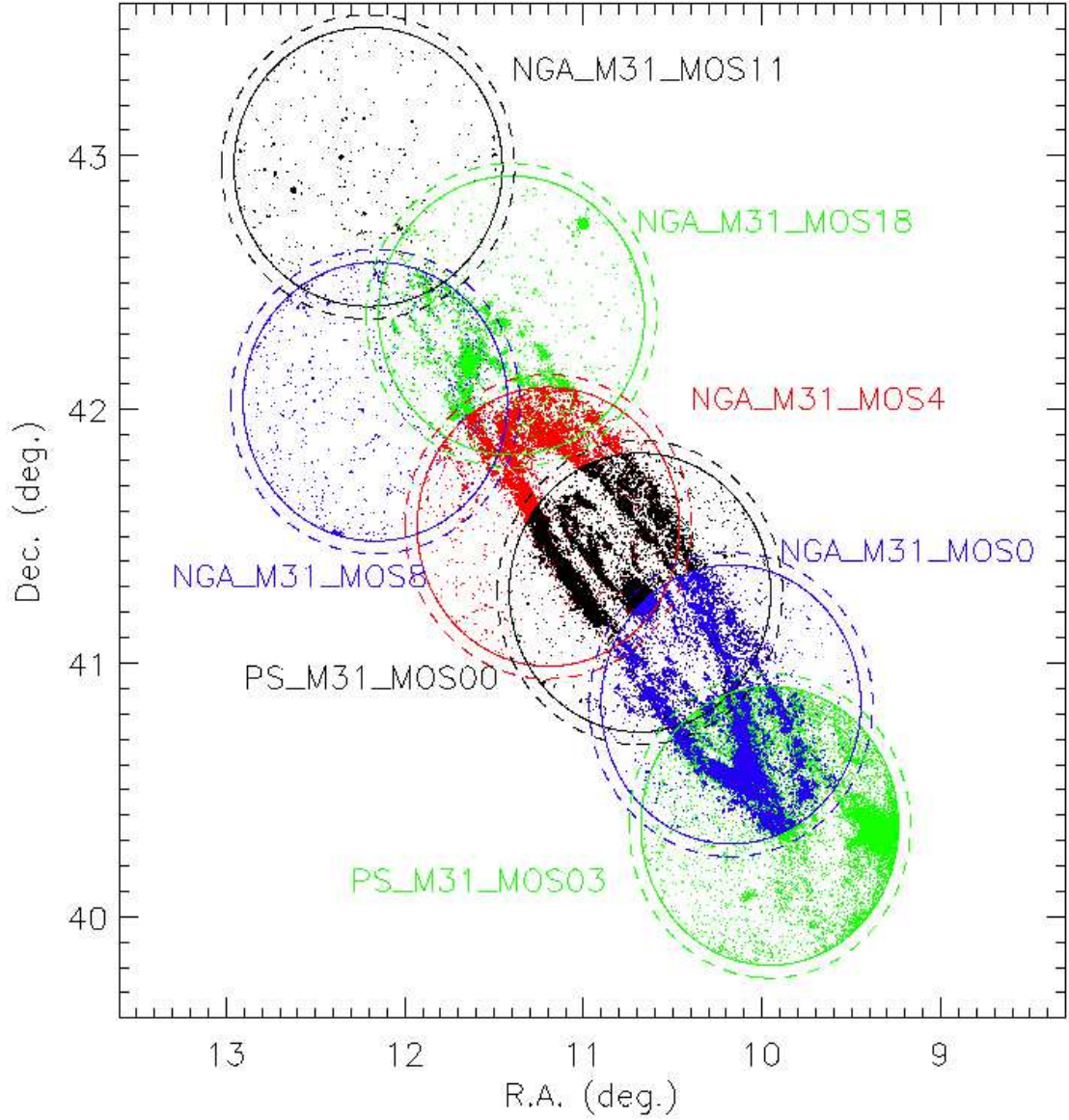} 
\end{center}
\caption{The detected FUV pixel map of the selected 7 $GALEX$ fields.
The solid circles are the FOV of our selection (1.1 degree diameter) and the dashed circles are the 
whole FOV of $GALEX$.\label{fig4}}
\end{figure}

\begin{figure} 
\begin{center}
\includegraphics[width=126mm]{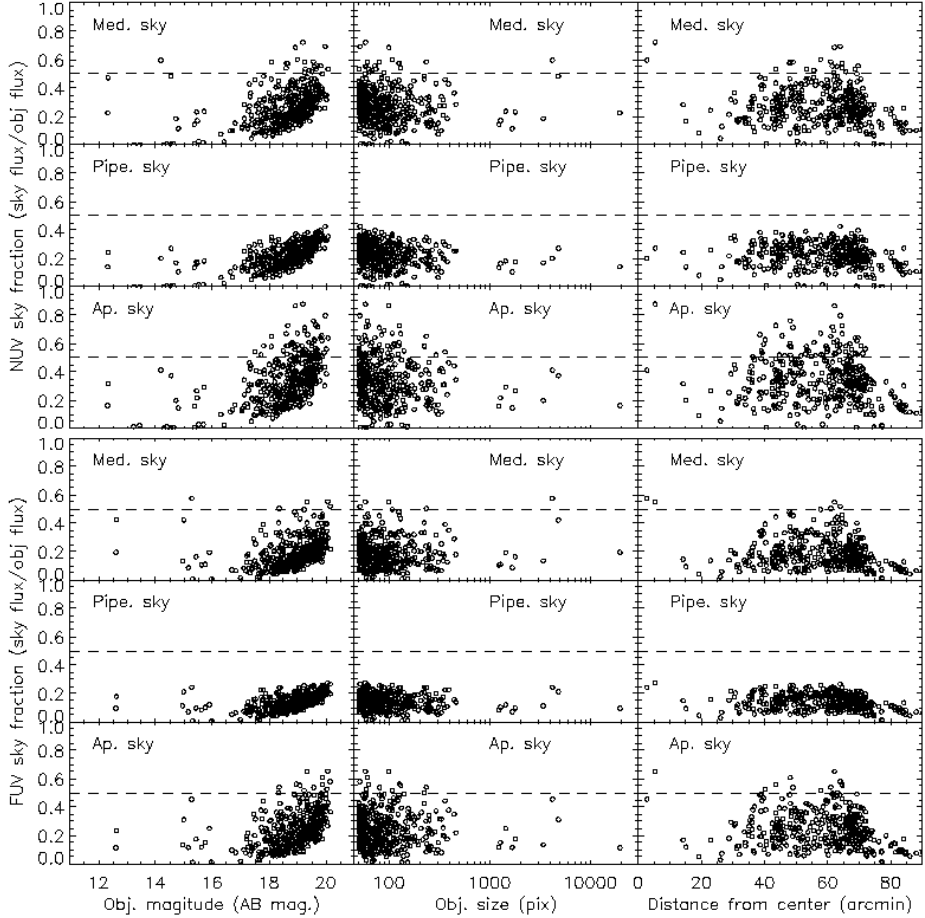} 
\end{center}
\caption{Comparison of three different methods of sky background estimations. 
The sky fraction in our photometry (``NGA\_M31\_MOS0" field) is displayed as the fraction of sky flux over 
object's original flux (sky un-subtracted flux). 
%%In the first column, it is plotted versus the measured magnitude of objects, in the second column versus 
In the first and second column, it is plotted with the measured magnitude of objects and 
%%the size of the objects. 
the size of the objects, respectively. 
In the third column, it is displayed with de-projected distance from the center of M31. 
The upper 9 panels are NUV measurements and the lower 9 panels are FUV. 
The background is always higher in NUV because this filter includes light from older, more diffuse populations. 
Therefore, the background subtraction is more critical for NUV.\label{fig5}}
\end{figure}

\begin{figure} 
\begin{center}
\includegraphics[width=166mm]{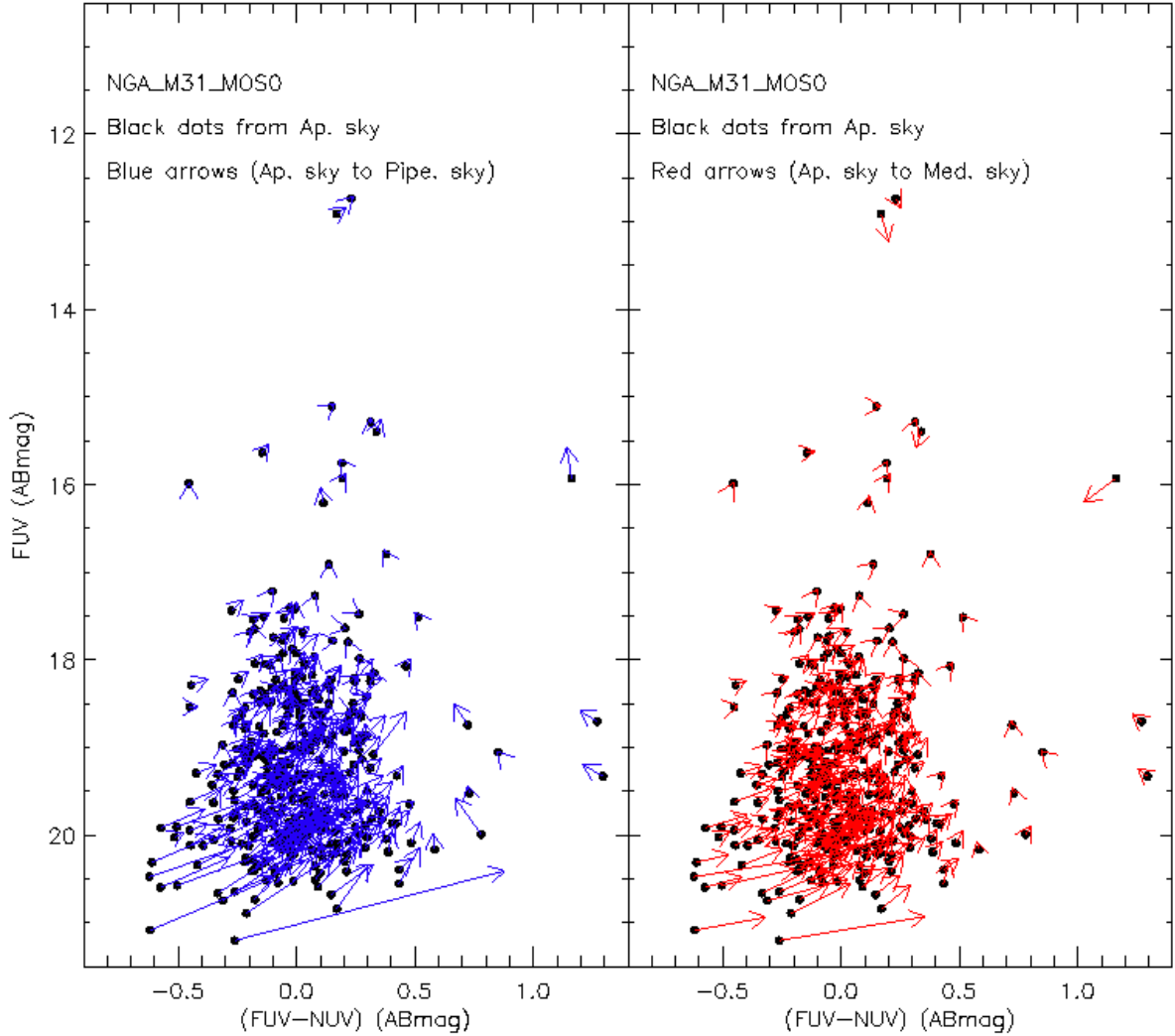} 
\end{center}
\caption{Comparison of photometry of the UV sources obtained from three different types of sky subtraction 
for the ``NGA\_M31\_MOS0" field. 
Black dots are obtained subtracting the sky flux measured from the local background. 
The end points of the blue arrows (left panel) are the photometry results using the GR4 pipeline sky. 
The end points of the red arrows (right panel) are from our median (3$\times$3 pixel) filtered sky.\label{fig6}}
\end{figure}

\begin{figure}
\begin{center}
\includegraphics[width=150mm]{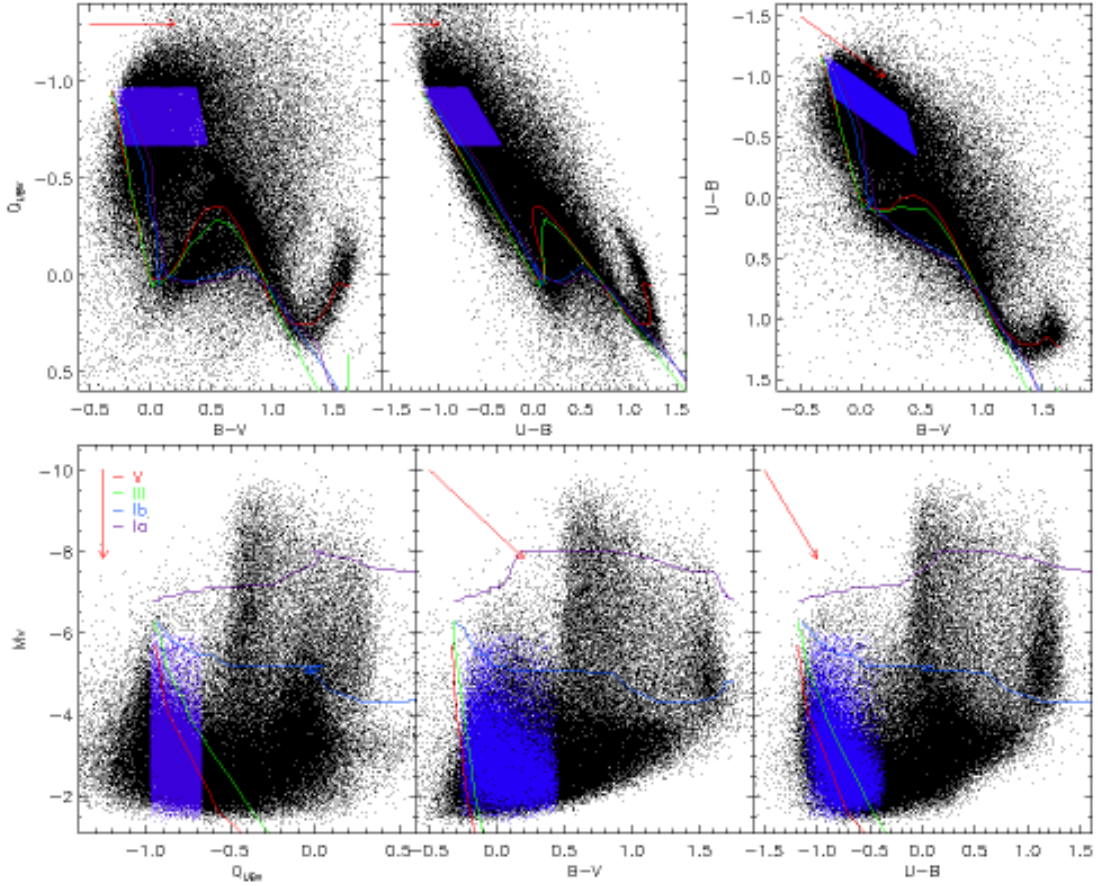}
\end{center}
\caption{The selection of OB type stars from the ground-based photometry of \cite{mas06}. 
The plotted points have magnitude error in $UBV$ bands lower than 0.1~mag. 
Blue points are our selected OB stars. 
The lines represent the intrinsic color as a function of T$_{eff}$ for luminosity classes main-sequence (V, red), 
giant (III, green), and supergiant (Ib and Ia, blue and purple) from \cite{all82}. 
%%The red arrow in each panel is the direction of reddening ($E(B-V) = 1.2$). \label{fig7}}
The red arrow in each panel is the direction of reddening with $E(B-V) = 0.7$. \label{fig7}}
%The accumulation of points along 
%some straight lines which appear as artifacts, are in fact eliminated by applying much more stringent 
%error cuts than we used. 
%However, our chosen error cuts ($<$ 0.1~mag) ensure good selection of the hottest stars which are used in our 
%analysis, therefore we did not need to further restrict the sample for our purposes.\label{fig7}}
\end{figure}

\begin{figure}
\begin{center}
\includegraphics[width=166mm]{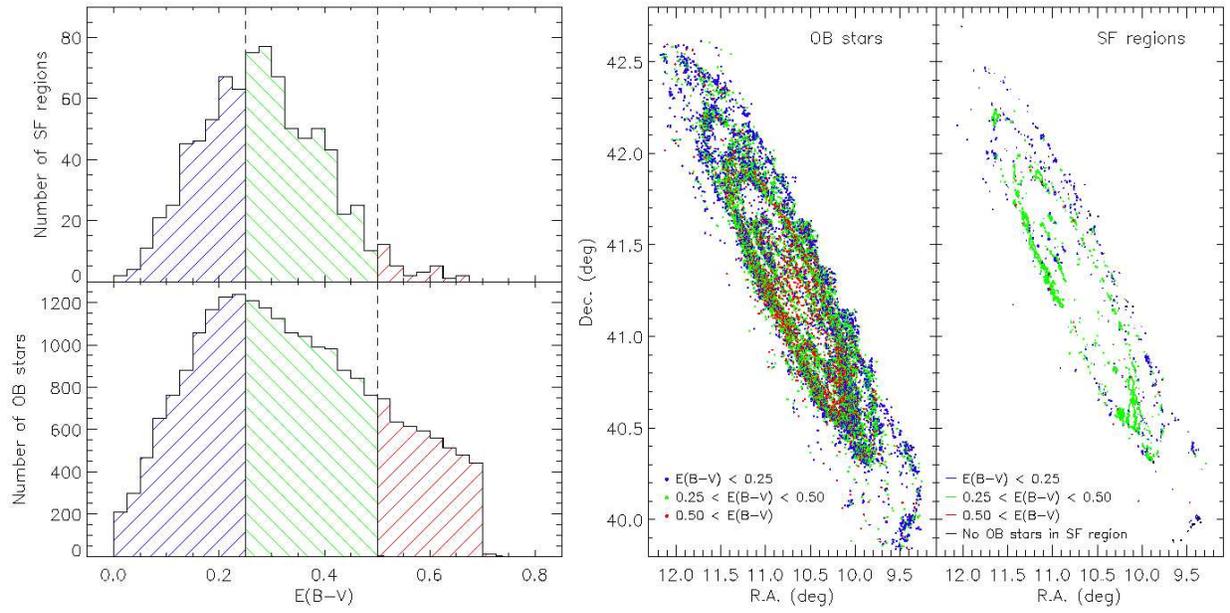}
\end{center}
\caption{Number distributions of reddening for the selected OB stars and for the detected SF regions 
(left panels). 
Spatial distributions of OB stars and SF regions are presented color-coded by three ranges of reddening 
in the right panels.\label{fig8}}
\end{figure}

\begin{figure}
\begin{center}
\includegraphics[width=83mm]{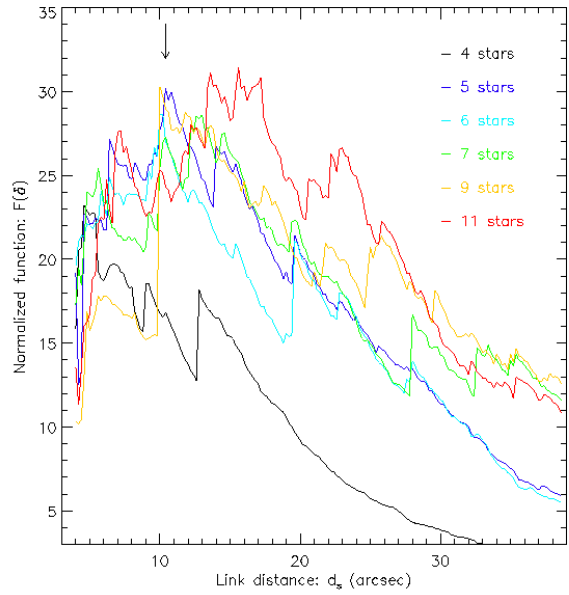}
\end{center}
\caption{The mean normalized fluctuation function changes with the minimum number of 
stars of an association (N$_{min}$) and the maximum link distance (d$_{s}$) between stars. 
The black arrow indicates the maximum peak of this function, which defined the parameters 
adopted for our selection of OB associations.\label{fig9}}
\end{figure}

\begin{figure}
\begin{center}
\includegraphics[width=166mm]{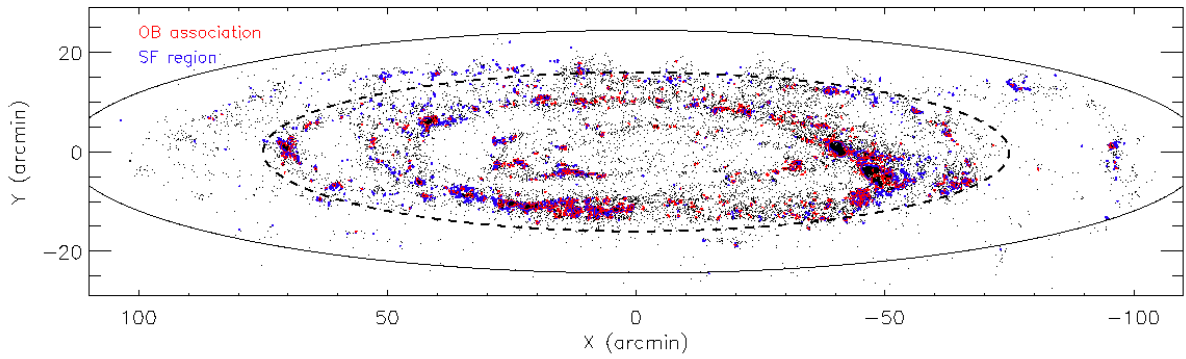}
\end{center}
\caption{Spatial distribution of SF regions detected from the $GALEX$ imaging (blue contours) 
and OB associations from optical stellar photometry (red contours). 
Black dots are the selected OB stars. 
The solid ellipse has a 26~kpc de-projected radius and the dashed ellipse has a 17~kpc de-projected 
radius.\label{fig10}}
\end{figure}

\begin{figure}
\begin{center}
\includegraphics[width=136mm]{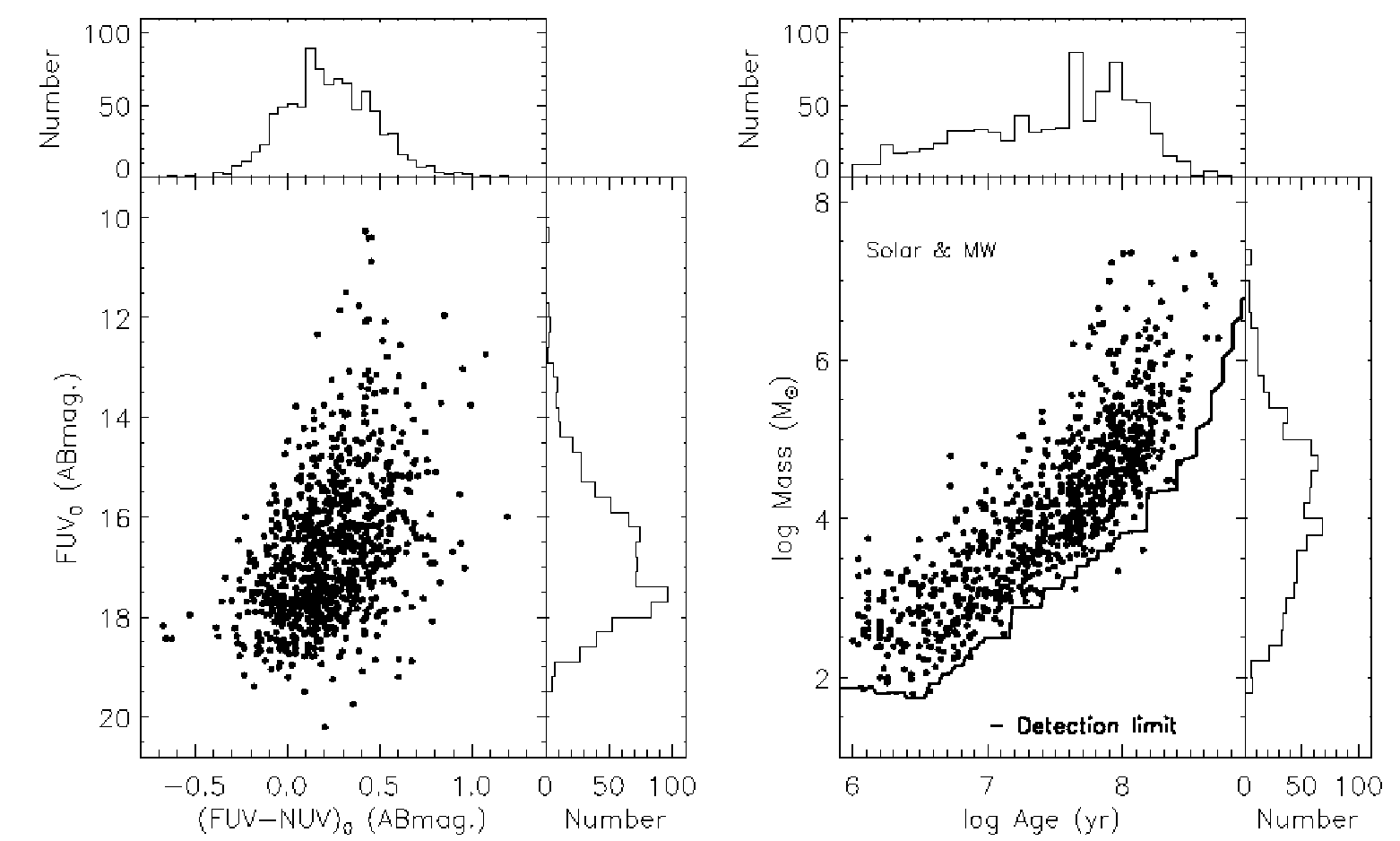}
\includegraphics[width=136mm]{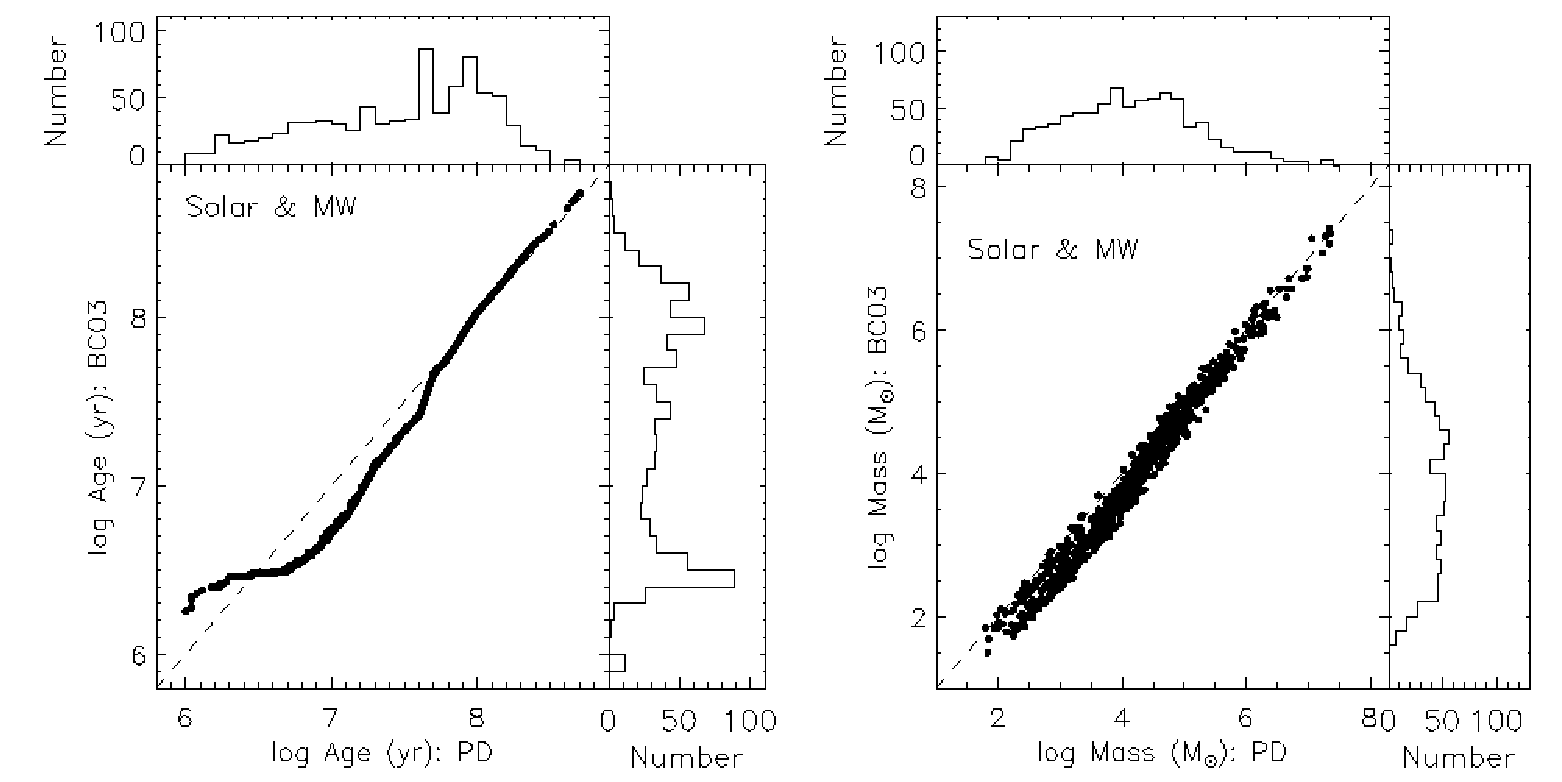}
\end{center}
\caption{Upper left panel: the colour-magnitude diagram of the SF regions. 
Upper right panel: the estimated ages and masses assuming solar metallicity (Z=0.02) 
with interstellar extinction of MW dust type using the PD model grid. 
The line below the data points is the model-estimate of our detection limit, 
based on the limiting magnitudes. 
Lower left/right panels: age/mass differences between two sets of models, assuming solar metallicity 
with MW dust type.\label{fig11}}
\end{figure}

\begin{figure}
\begin{center}
\includegraphics[width=146mm]{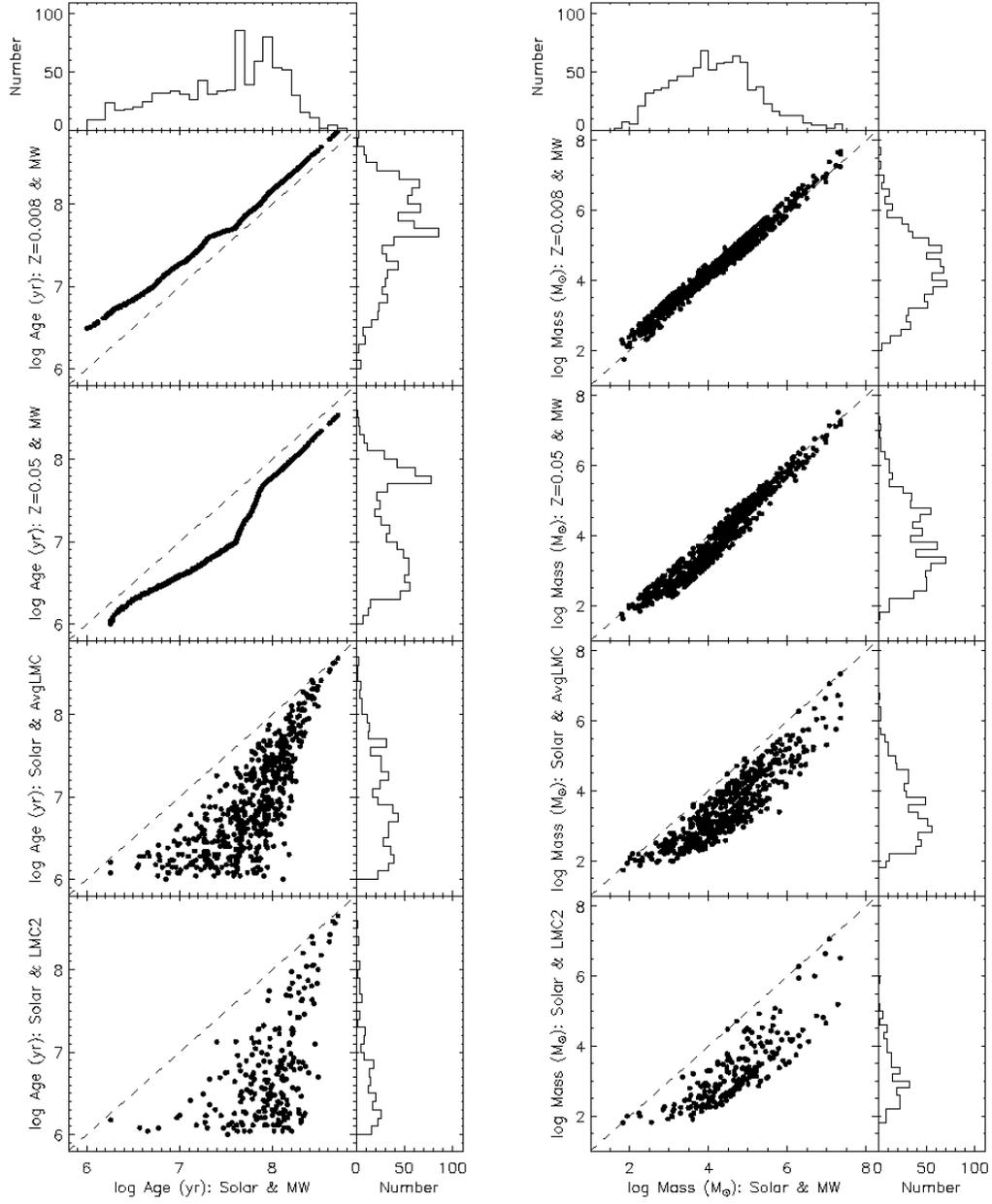}
\end{center}
\caption{Ages and masses of the SF regions, derived assuming various metallicities and dust types. 
Results from the PD models are shown.\label{fig12}}
\end{figure}

\begin{figure}
\begin{center}
\includegraphics[width=166mm]{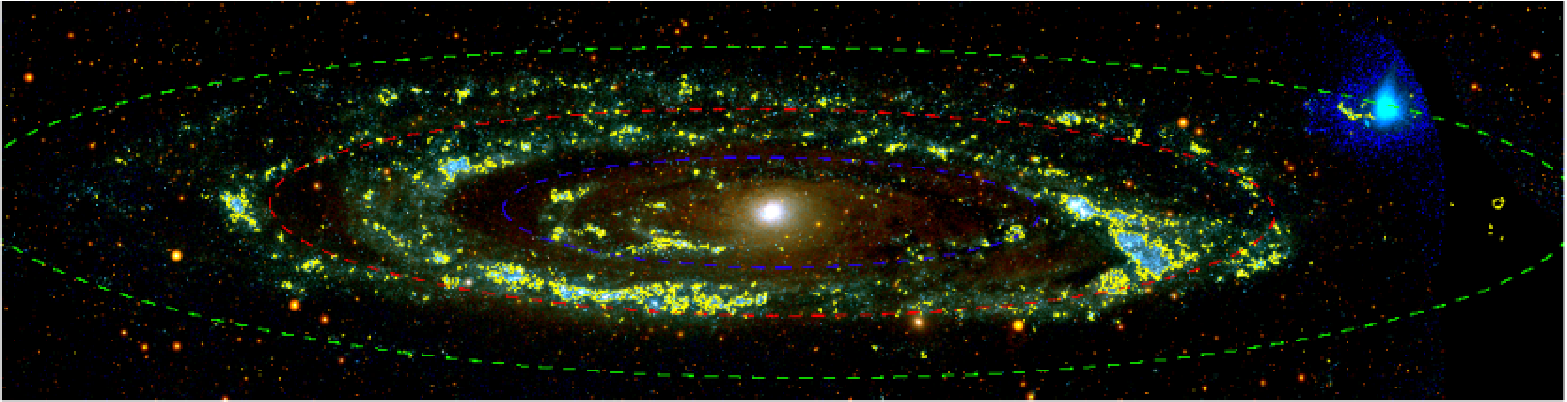}
\includegraphics[width=166mm]{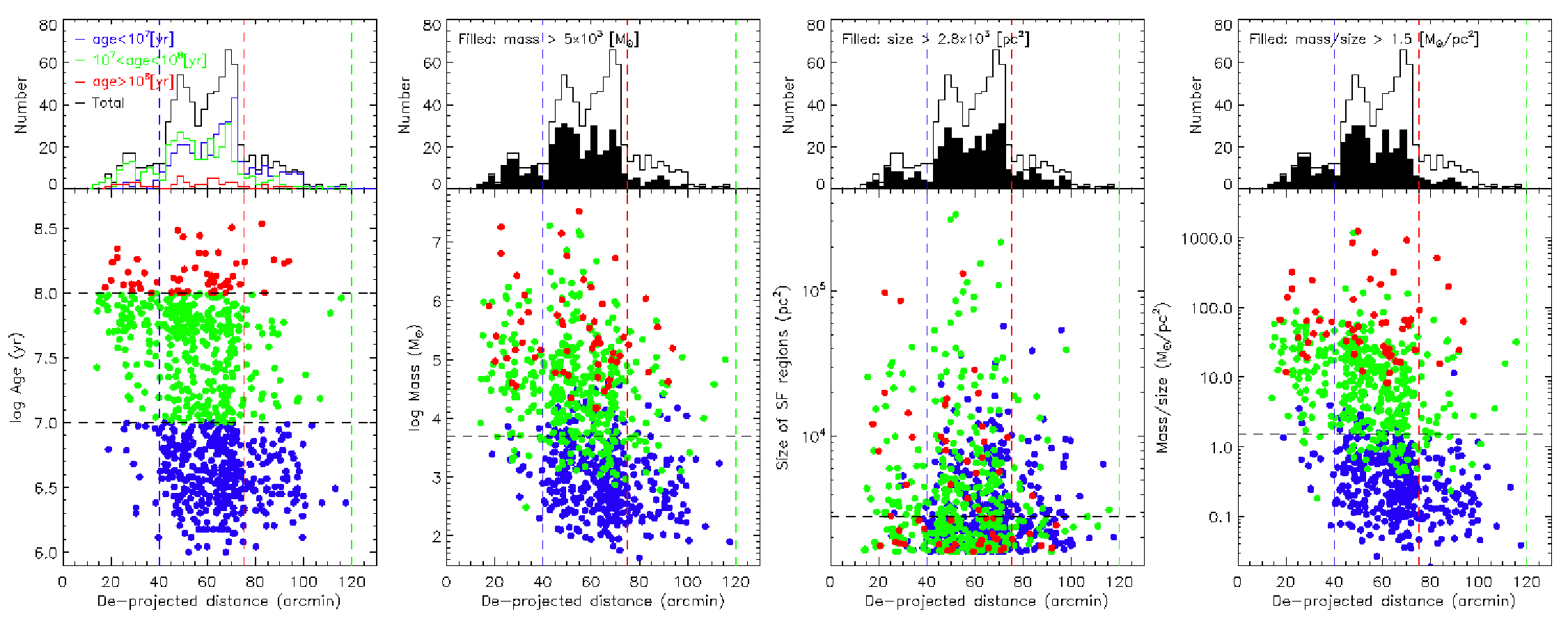}
\end{center}
\caption{Top panel: the spatial distribution of SF regions (yellow contours) 
on a colour composite image (blue: FUV, green: FUV+NUV, red: NUV). 
%%
%%The blue and bright blob on the outermost ellipse is a bright foreground star (HD 3431). 
The blue and bright blob on the outermost ellipse is a bright Galactic foreground star (HD 3431). 
Bottom panels: distributions of ages, masses, size, and mass-per-unit-area of SF regions against the 
de-projected distance from the center of M31. 
Different symbol colors indicate different age bins (red: older than 100~Myrs, green: 10 - 100~Myrs, 
blue: younger than 10~Myrs). 
The dashed vertical lines correspond to the ellipses drawn on the top image at 40, 75, and 120~arcmin, 
%%corresponding to deprojected distances in M31 9, 17, and 27~kpc, respectively.
corresponding to deprojected distances in M31 of 9, 17, and 27~kpc, respectively.
%The dashed vertical lines correspond to the ellipses drawn on the top image at 18, 28, 52, 68, 83, 98, and 120~arcmin 
%(4, 6, 11, 16, 19, 22, and 27~kpc), respectively. 
The ages and masses shown here were derived using metallicity Z=0.05 and MW (R$_V$=3.1) dust type.\label{fig13}}
\end{figure}

\begin{figure}
\begin{center}
\includegraphics[width=83mm]{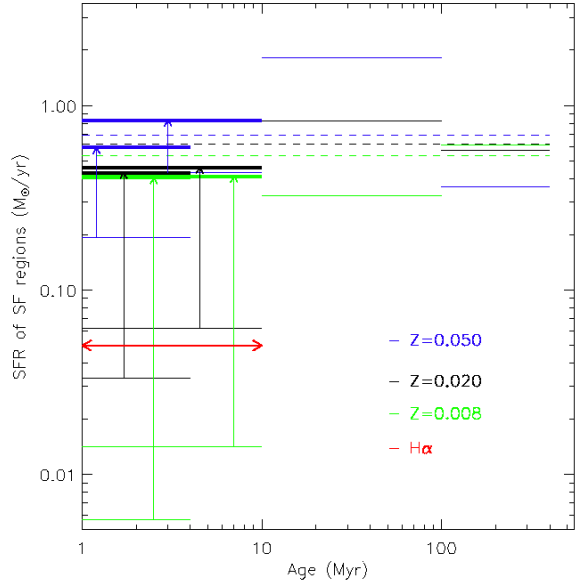}
\end{center}
%%\caption{Star Formation Rates at recent epochs in M31, from our SF regions. 
\caption{SFRs at recent epochs in M31, from our SF regions. 
Black lines are values derived for solar metallicity, blue lines are for Z=0.05, and green lines
are for Z=0.008.
In all cases MW-type dust was assumed to correct the UV luminosities for interstellar extinction,
as discussed in Section 6.
Solid horizontal lines represent the SFR in four time bins: $<$4, $<$10, 10-100, and 100-400~Myrs,
and the dashed lines are mean values over the past 400~Myrs.
The red line is the SFR from H$\alpha$ \citep{mas07}, shown in the $<$10~Myrs
age bin which traces only the youngest stars capable of ionizing the ISM.
The thick horizontal lines above vertical arrows on the $<$4~Myrs age bin and $<$10~Myrs indicate 
the total SFRs obtained by adding the SFR from IR measurements \citep{bar06} to our UV-based estimates.
\label{fig14}}
\end{figure}

\clearpage

%%%%%%%%%%Table%%%%%%%%%%%%%%%%%%%%%

%%Table1
\begin{deluxetable}{rrrrrrr}
\tablecolumns{7} 
\tablewidth{0pc}
\tablecaption{Details of the selected $GALEX$ fields of M31's disk region} 
\tablehead{
%%\colhead{N.} & \colhead{Field Name} & \colhead{R.A.} & \colhead{Dec.} & \colhead{Exp. time} & \colhead{Adopted thresh.} & \colhead{Detected pixels}\\
\colhead{No} & \colhead{Field Name} & \colhead{R.A.} & \colhead{Dec.} & \colhead{Exp. time} & \colhead{Adopted thresh.} & \colhead{Detected pixels}\\
\colhead{} & \colhead{} & \colhead{[deg]} & \colhead{[deg]} & \colhead{[s]} & \colhead{[c/s]} & \colhead{above thresh.}
}
\startdata
%\tableline
1 & NGA\_M31\_MOS11 & 12.204914 & 42.956084 & 3340.35 & 0.00397 &   5533\\ 
2 & NGA\_M31\_MOS8  & 12.164794 & 42.030947 & 2702.75 & 0.00384 &  36791\\ 
3 & NGA\_M31\_MOS18 & 11.403102 & 42.370208 & 3244.45 & 0.00365 &  77217\\
4 & NGA\_M31\_MOS4  & 11.253077 & 41.863775 & 3589.70 & 0.00373 & 209395\\ 
5 & PS\_M31\_MOS00  & 10.683594 & 41.277741 & 3760.10 & 0.00353 & 188512\\
6 & NGA\_M31\_MOS0  & 10.173990 & 40.836523 & 6811.30 & 0.00324 & 188847\\
7 & PS\_M31\_MOS03  &  9.957439 & 40.358496 & 1182.20 & 0.00446 & 152669\\
\enddata
\end{deluxetable}

%%Table2
\begin{center}
\begin{deluxetable}{rrrrrrrrrrrrrrrrr}
\rotate
\setlength{\tabcolsep}{0.03in}
\tabletypesize{\tiny}
\tablecolumns{17} 
\tablewidth{0pc}
\tablecaption{The UV detected SF regions in M31.\tablenotemark{a}} 
\tablehead{
\colhead{Id} & \colhead{R.A.$_{J2000}$} & \colhead{Dec.$_{J2000}$} & \colhead{FUV} & \colhead{FUV$_{err}$} & \colhead{NUV} & \colhead{NUV$_{err}$} & \colhead{$E(B-V)$} & \colhead{Area}
& \colhead{Age\tablenotemark{b,c}} & \colhead{Age$_{min/max}$\tablenotemark{b,c}} & \colhead{Mass\tablenotemark{b,c}} & \colhead{Mass$_{min/max}$\tablenotemark{b,c}}
& \colhead{Age\tablenotemark{b,d}} & \colhead{Age$_{min/max}$\tablenotemark{b,d}} & \colhead{Mass\tablenotemark{b,d}} & \colhead{Mass$_{min/max}$\tablenotemark{b,d}}\\
\colhead{} & \colhead{[deg]} & \colhead{[deg]} & \colhead{[ABmag]} & \colhead{[ABmag]} & \colhead{[ABmag]} & \colhead{[ABmag]} & \colhead{[mag]} & \colhead{[arcsec$^2$]}
& \colhead{[Myrs]} & \colhead{[Myrs]/[Myrs]} & \colhead{[M$_\sun$]} & \colhead{[M$_\sun$]/[M$_\sun$]}
& \colhead{[Myrs]} & \colhead{[Myrs]/[Myrs]} & \colhead{[M$_\sun$]} & \colhead{[M$_\sun$]/[M$_\sun$]}}
\startdata
%\tableline
1003  & 12.013985  & 42.690975  & 18.870  & 0.020  & 19.158  & 0.013  & 0.20  &    112.5  &   1.6  &   1.3/  1.9  &  3.7e+02  &  4.4e+02/ 3.7e+02  & -99.0  & -99.0/  1.2  & -9.9e+01  & -9.9e+01/ 3.8e+02 \\
2040  & 11.951906  & 42.073174  & 19.124  & 0.025  & 19.331  & 0.016  & 0.06  &    126.0  &   4.1  &   2.9/  5.7  &  1.4e+02  &  9.0e+01/ 2.2e+02  &   2.3  &   1.8/  2.7  &  8.2e+01  &  8.2e+01/ 1.1e+02 \\
3069  & 11.725769  & 41.968216  & 19.770  & 0.034  & 19.986  & 0.023  & 0.05  &    155.2  &   3.7  &   2.1/  5.8  &  6.8e+01  &  5.4e+01/ 1.1e+02  &   2.1  &   1.4/  2.7  &  4.2e+01  &  5.6e+01/ 5.6e+01 \\
3071  & 11.715304  & 41.971684  & 18.494  & 0.018  & 18.420  & 0.010  & 0.13  &    328.5  &  58.7  &  49.9/ 66.5  &  1.5e+04  &  1.2e+04/ 1.9e+04  &  22.4  &  17.6/ 30.8  &  5.0e+03  &  5.0e+03/ 8.2e+03 \\
3074  & 11.641906  & 41.971554  & 20.070  & 0.039  & 20.057  & 0.023  & 0.19  &    117.0  &  40.5  &  24.5/ 49.7  &  3.2e+03  &  1.4e+03/ 4.4e+03  &  10.1  &   6.8/ 17.5  &  7.4e+02  &  4.5e+02/ 1.8e+03 \\
3077  & 11.618323  & 41.983509  & 18.498  & 0.019  & 18.633  & 0.012  & 0.30  &    452.2  &   7.1  &   5.7/  8.8  &  3.3e+03  &  2.5e+03/ 4.9e+03  &   3.1  &   2.8/  3.6  &  1.2e+03  &  1.2e+03/ 1.8e+03 \\
3081  & 11.635375  & 41.991177  & 16.531  & 0.007  & 16.393  & 0.004  & 0.19  &   2072.2  &  79.4  &  76.8/ 82.0  &  2.2e+05  &  2.2e+05/ 2.2e+05  &  48.0  &  45.3/ 50.4  &  1.6e+05  &  1.6e+05/ 1.6e+05 \\
3082  & 11.658252  & 41.985649  & 18.267  & 0.016  & 18.131  & 0.008  & 0.26  &    229.5  &  74.8  &  68.9/ 80.6  &  6.4e+04  &  6.4e+04/ 7.6e+04  &  43.3  &  34.6/ 49.3  &  4.2e+04  &  2.8e+04/ 5.6e+04 \\
3083  & 11.625877  & 41.988590  & 19.501  & 0.032  & 19.476  & 0.019  & 0.46  &    207.0  &  19.9  &  14.8/ 29.8  &  1.8e+04  &  7.8e+03/ 3.0e+04  &   6.2  &   4.8/  8.1  &  4.0e+03  &  2.6e+03/ 6.8e+03 \\
3085  & 11.648593  & 41.992626  & 20.509  & 0.055  & 20.507  & 0.033  & 0.30  &    112.5  &  28.4  &  15.2/ 47.0  &  3.6e+03  &  2.2e+03/ 6.8e+03  &   7.7  &   4.8/ 15.5  &  8.1e+02  &  3.1e+02/ 2.7e+03 \\
3086  & 11.654299  & 41.999249  & 17.305  & 0.011  & 17.136  & 0.006  & 0.20  &   1055.2  &  89.3  &  85.2/ 94.1  &  1.4e+05  &  1.4e+05/ 1.4e+05  &  54.9  &  52.2/ 57.6  &  8.9e+04  &  8.9e+04/ 1.2e+05 \\
 ...  &       ...  &       ...  &   ...   &  ...   &   ...   &  ...   &  ...  &      ...  &   ...  &    .../...   &       ... &         .../...    &    ... &    .../...   &       ... &        .../...    \\

\enddata
\tablenotetext{a}{Full catalog available in electronic version.}
\tablenotetext{b}{``-99.0" and ``-9.9e+01" indicate the cases where the observed color is outside the range of model colors at all ages.}
\tablenotetext{c}{Metallicity Z=0.02 and average MW ($R_V$ = 3.1) dust type.}
\tablenotetext{d}{Metallicity Z=0.05 and average MW ($R_V$ = 3.1) dust type.}
\end{deluxetable}
\end{center}

%%Table3
\begin{deluxetable}{cccccc}
%\tablecolumns{7} 
\tablewidth{0pc}
\tablecaption{M31 SFR derived from UV flux in different age intervals.} 
\tablehead{
\colhead{Metallicity} &\multicolumn{5}{c}{SFR[$M_{\sun}$/yr] from UV-detected SF regions}\\ \cline{2-6}
\colhead{(Z)} & \colhead{$<$4~Myrs\tablenotemark{*}} & \colhead{$<$10~Myrs\tablenotemark{*}} & \colhead{10-100~Myrs} & \colhead{100-400~Myrs} & \colhead{$<$400~Myrs}
}
\startdata
0.008 & 0.006(0.406) & 0.014(0.414) & 0.325 & 0.610 & 0.532\\
0.020 & 0.033(0.433) & 0.062(0.462) & 0.826 & 0.569 & 0.616\\
0.050 & 0.192(0.592) & 0.434(0.834) & 1.811 & 0.361 & 0.690\\
%0.008 & 0.006(0.406) & 0.037(0.437) & 0.427 & 0.775 & 0.680\\ 
%0.020 & 0.038(0.438) & 0.102(0.502) & 1.473 & 0.629 & 0.807\\ 
%0.050 & 0.329(0.729) & 0.599(0.999) & 2.259 & 0.494 & 0.894\\ 
\enddata
\tablenotetext{*}{The value inside parentheses is obtained by adding the SFR from IR measurements \citep{bar06} to the SFR derived from our UV measurements.}
\end{deluxetable}


\begin{thebibliography}{}

\bibitem[Aller et al.(1982)]{all82} Aller, L.~H., et al.\ 1982, Landolt-Bornstein: Numerical Data and Functional Relationships in Science and Technology

\bibitem[Barmby et al.(2006)]{bar06} Barmby, P., et al.\ 2006, \apjl, 650, L45

\bibitem[Battinelli(1991)]{bat91} Battinelli, P.\ 1991, \aap, 244, 69

\bibitem[Bianchi(2006)]{bia07a} Bianchi, L., \& et al.\ 2006, The Ultraviolet Universe: Stars from Birth to Death,
26th meeting of the IAU, Joint Discussion 4, 16-17 August 2006, Prague, Czech Republic, JD04, 37, 4

\bibitem[Bianchi(2009)]{bia09} Bianchi, L.\ 2009, \apss, 320, 11

\bibitem[Bianchi \& Efremova(2006)]{bia06} Bianchi, L., \& Efremova, B.~V.\ 2006, \aj, 132, 378

\bibitem[Bianchi et al.(1996)]{bia96} Bianchi, L., Clayton, G.~C., Bohlin, R.~C., Hutchings, J.~B., \& Massey, P.\ 1996, \apj, 471, 203

\bibitem[Bianchi et al.(2003)]{bia03} Bianchi, L., Madore, B., Thilker, D., Gil de Paz, A., Martin, C., \& The GALEX Team\ 2003, The Local Group as an Astrophysical Laboratory, 10

\bibitem[Bianchi et al.(2007)]{bia07} Bianchi, L., et al.\ 2007, \apjs, 173, 659

\bibitem[Block et al.(2006)]{blo06} Block, D.~L., et al.\ 2006, \nat, 443, 832

%\bibitem[Bressan et al.(2002)]{bre02} Bressan, A., Silva, L., \& Granato, G.~L.\ 2002, \aap, 392, 377

\bibitem[Bruzual \& Charlot(2003)]{bru03} Bruzual, G., \& Charlot, S.\ 2003, \mnras, 344, 1000

\bibitem[Cardelli et al.(1989)]{car89} Cardelli, J.~A., Clayton, G.~C., \& Mathis, J.~S.\ 1989, \apj, 345, 245

\bibitem[Efremova \& Bianchi(2009)]{efr09} Efremova, B.~V. \& Bianchi, L.\ 2009, in preparation
 
\bibitem[Fuchs et al.(2009)]{fuc09} Fuchs, B., Jahrei{\ss}, H., \& Flynn, C.\ 2009, \aj, 137, 266 

\bibitem[Gordon \& Clayton(1998)]{gor98} Gordon, K.~D., \& Clayton, G.~C.\ 1998, \apj, 500, 816

\bibitem[Gordon et al.(2006)]{gor06} Gordon, K.~D., et al.\ 2006, \apjl, 638, L87 

\bibitem[Hammer et al.(2007)]{ham07} Hammer, F., Puech, M., Chemin, L., Flores, H., \& Lehnert, M.~D.\ 2007, \apj, 662, 322 

\bibitem[Hou et al.(2009)]{hou08} Hou, J., Yin, J., Boissier, S., Prantzos, N., Chang, R.~X., \& Chen, L.\ 2009, IAU Symposium, 254, 27 

%\bibitem[Ibata et al.(2001)]{iba01} Ibata, R., Irwin, M., Lewis, G., Ferguson, A.~M.~N., \& Tanvir, N.\ 2001, \nat, 412, 49

\bibitem[Ibata et al.(2007)]{iba07} Ibata, R., Martin, N.~F., Irwin, M., Chapman, S., Ferguson, A.~M.~N., Lewis, G.~F., \& McConnachie, A.~W.\ 2007, \apj, 671, 1591 

\bibitem[Ivanov(1996)]{iva96} Ivanov, G.~R.\ 1996, \aap, 305, 708

\bibitem[Ivanov(1998)]{iva98} ------.\ 1998, \aap, 337, 39

\bibitem[Kennicutt(1998)]{ken98} Kennicutt, R.~C., Jr.\ 1998, \apj, 498, 541 

\bibitem[Magnier et al.(1993)]{mag93} Magnier, E.~A., et al.\ 1993, \aap, 278, 36

\bibitem[Martin et al.(2005)]{mar05} Martin, D.~C., et al.\ 2005, \apjl, 619, L1 

\bibitem[Massey(2003)]{mas03} Massey, P.\ 2003, \araa, 41, 15 

\bibitem[Massey et al.(1995)]{mas95} Massey, P., Armandroff, T.~E., Pyke, R., Patel, K., \& Wilson, C.~D.\ 1995, \aj, 110, 2715

\bibitem[Massey et al.(2006)]{mas06} Massey, P., Olsen, K.~A.~G., Hodge, P.~W., Strong, S.~B., Jacoby, G.~H., Schlingman, W., \& Smith, R.~C.\ 2006, \aj, 131, 2478

\bibitem[Massey et al.(2007)]{mas07} Massey, P., Olsen, K.~A.~G., Hodge, P.~W., Jacoby, G.~H., McNeill, R.~T., Smith, R.~C., \& Strong, S.~B.\ 2007, \aj, 133, 2393

\bibitem[McConnachie et al.(2005)]{mcc05} McConnachie, A.~W., Irwin, M.~J., Ferguson, A.~M.~N., Ibata, R.~A., Lewis, G.~F., \& Tanvir, N.\ 2005, \mnras, 356, 979

\bibitem[Misselt et al.(1999)]{mis99} Misselt, K.~A., Clayton, G.~C., \& Gordon, K.~D.\ 1999, \apj, 515, 128

\bibitem[Morrissey et al.(2007)]{mor07} Morrissey, P., et al.\ 2007, \apjs, 173, 682

%\bibitem[Oke \& Gunn(1983)]{oke83} Oke, J.~B., \& Gunn, J.~E.\ 1983, \apj, 266, 713 

%\bibitem[Schlegel et al.(1998)]{sch98} Schlegel, D.~J., Finkbeiner, D.~P., \& Davis, M.\ 1998, \apj, 500, 525

\bibitem[Tolea(2009)]{tol09} Tolea, A.~C.\ 2009, Ph.D.~Thesis

\bibitem[Walterbos \& Kennicutt(1988)]{wal88} Walterbos, R.~A.~M., \& Kennicutt, R.~C., Jr.\ 1988, \aap, 198, 61

\end{thebibliography}
\end{document}